\documentstyle[preprint,eqsecnum,aps]{revtex}

\begin{document}
\draft

\preprint{HD--TVP--96--04}

\title{\vspace*{1cm} Elastic Scattering and Transport Coefficients for
       a Quark Plasma in $SU_f(3)$ at Finite Temperatures}

\author{P.~Rehberg\thanks{E-Mail:
        {\tt Peter@Frodo.TPhys.Uni-Heidelberg.DE}},
        S.~P.~Klevansky and J.~H\"ufner}
\address{Institut f\"ur Theoretische Physik, Universit\"at Heidelberg, \\
         Philosophenweg 19, D--69120 Heidelberg, Germany}

\date{April 1996}

\maketitle
\vspace{1cm}
\begin{abstract}
The temperature dependence of the elastic scattering processes $qq'\to
qq'$ and $q\bar q'\to q\bar q'$, with $q, q' = u, d, s$ is studied as a
function of the scattering angle and the center of mass energy of the
collision within the framework of the $SU_f(3)$ Nambu--Jona--Lasinio
model.  Critical scattering at threshold is observed in the $q\bar
q'\to q\bar q'$ process, leading to an enhancement of the cross section
as occurs in the phenomenon of critical opalescence. Transport
properties such as viscosity, mean free paths and thermal relaxation
times are calculated. Strangeness enhancement is investigated via the
chemical relaxation times, which are found to be considerably higher
than those calculated via perturbative QCD. A comparison with the
experimental values for the strangeness enhancement in $S+S$ collisions
leads to an upper limit of 4~fm/$c$ for the lifetime of the plasma. \\[3mm]
PACS numbers: 11.30.Rd, 12.38.Mh, 24.85.+p, 11.10.Wx
\end{abstract}

\section{Introduction} \label{introsec}
With the proposed experiments for heavy ion collisions at the LHC
coming closer, the onus has settled on the theoretical physics
community to provide a clear, coherent and credible description of such
collisions. One of the major goals, that of detecting the presence
of plasma formation of the constituent quarks and gluons is for both
theorists and experimentalists alike, a formidable task.   Such a phase
is believed to be formed as matter is deconfined.

From the theoretical side, a description that goes beyond the standard
equilibrium statistical approaches is necessitated by the short collision
time of the ions, since this may preclude equilibration of all degrees
of freedom. Thus a transport theory that is consistently based on
quantum chromodynamics (QCD) itself is required.   While such
formulations have been set up by some authors \cite{heinz,elze}, they
have not reached the stage where they can be implemented to provide
original physical results.

One step in this direction is to examine the consequences of a transport
theory based on an effective model of QCD, that exhibits some important 
features of the theory. In particular, it is desirable to see if any
major effects due to the exhibited feature are observable in the final
analysis.

It is well-known that the low energy particle sector is well described
by effective chiral theories of QCD. We therefore choose to implement
the Nambu--Jona-Lasinio (NJL) model \cite{vogl,sandi,hatkun} in its
three flavor extension for such a study. This model has the advantage
that, being a model, it can be studied over the entire temperature
range. The NJL model also offers a simple intuitive view of chiral
symmetry breakdown and restoration via a realization of the pairing
mechanism of quark and antiquark states of an effective interaction in
a similar sense to the BCS theory of superconductivity. The model
suffers however from several unaesthetic diseases: the simple form of a
point-like interaction, while making the model analytically tractable,
also ensures its nonrenomalizability, and a cutoff scale $\Lambda$ for
the theory must be introduced. A confinement / deconfinement phase
transition cannot occur. Rather a ``soft'' form of deconfinement is
observed at the so-called Mott temperature, at which the mesons
dissociate into their constituents \cite{gerry}.

The NJL model, being a strong coupling field theory, cannot be handled
in perturbation theory. The by now standard technique is a
self-consistent expansion of quantities to be calculated in the inverse
number of colors $1/N_c$ \cite{ncexpand}. It is this expansion, taken
to lowest order, that we use in our model.

In order to make some progress in the ambitious task presented, we have
already studied hadronization processes of quarks and antiquarks into
two mesons, $q\bar q '\to MM'$ \cite{su3hadron}, within the $SU_f(3)$
NJL model. In addition, we require the elastic scattering cross
sections for both quarks and quarks $q q'\to q q'$ and quarks
and antiquarks $q\bar q'\to q \bar q'$ within the three flavor
model. The cross sections for the third possible class, $\bar q\bar
q'\to\bar q\bar q'$, follow immediately from $q q'\to q q'$ by time
reversal invariance, as long as no chemical potential is involved.  It
is the purpose of this paper to calculate these cross sections,
together with the attendant transport properties (mean free paths and
shear viscosities) in the $SU_f(3)$ model for realistic values of the current
quark masses $m_{0u}=m_{0d}=5.5~{\rm MeV}$, $m_{0s}=140~{\rm MeV}$. We
note that some elastic scattering cross sections have been calculated
in the two flavor sector with the additional restriction of the chiral
limit condition $m_{0u}=m_{0d}=0$ in Ref.~\cite{su2elast}. As will be
seen, the extension presented here indicates significant increases in
the cross sections due to additional available channels.

The quark quark scattering cross sections are relatively featureless.
On the other hand, quark antiquark scattering shows a threshold
divergence at the Mott temperature $T_{M\pi}$, at which the pion
dissociates into its constituents. At higher temperatures that are to
be expected in the plasma phase, a large resonant structure is
observed, that is a remnant of this feature of critical scattering.
This is however not directly observable. When these cross
sections are included in a fully consistent transport theory based on
this chiral model of QCD, it is hoped that this may lead to observable
effects.

We have also studied transport coefficients witin this model.  We find
that the viscosity coefficient $\eta/T^3$ lies in the range 0.83 to 1.0
per flavor degree of freedom for light quarks, and is slightly lower
for strange quarks, being 0.68 to 0.75, i.e. the {\em total} viscosity
is found to lie between 3.2 and 4.0 considering only two flavors, while
for three flavors the total viscosity would lie between 4.6 and 5.5.
Perturbative QCD estimates lie slightly higher than this value: a
variational calculation \cite{baym} (see also \cite{hoka}) gives
$\eta/T^3\approx -1.16/(\alpha_s^2\log\alpha_s)$ for two quark flavors,
where $\alpha_s$ is the strong coupling strength. One finds the value
$\eta/T^3\approx6.3$ at $\alpha_s=0.6$. This should be viewed somewhat
critically, since this perturbative formula is strictly valid for
$\alpha_s\ll 1$, and is an extremely sensitive function of $\alpha_s$.
In the strongly interacting non--perturbative regime, it is not
applicable. Hydrodynamical estimates, on the other hand \cite{danigu},
give a wide range for this quantity, $2<\eta/T^3<3\tau T$, where $\tau$
is the proper time for the expansion of the system. We also calculate
mean free paths for light and heavy flavors and these are found to be
of the order of magnitude of 1~fm. The thermal relaxation times are
found to vary between 1--2~fm/$c$, with that of the strange quark being
slightly higher.

In order to discuss strangeness enhancement, we examine the chemical
relaxation time for strange quarks within the model. These, lying
between 11~fm/$c$ and 17~fm/$c$, are considerably higher than the naive
perturbative QCD prediction of Ref.~\cite{KMR} that proceeds via gluon
fusion. The origin of this discrepancy can be traced back to the value
of the strange quark mass $m_s$.  In the perturbative QCD calculation
of Ref.~\cite{KMR}, the physical mass was assumed to be equal to the
current quark mass value. In the NJL model, on the other hand, a
significantly enhanced value of the strange quark mass is still found
at temperatures of the order of 250~MeV \cite{su3hadron}. In fact it is
at least twice the current quark mass value. A recalculation of the
perturbative QCD calculation using the temperature dependent NJL masses
gives relaxation times of the same order of magnitude as our
calculation. We note that an assessment of the relevance of quark
antiquark scattering for the strangeness enhancement that is observed
e.g. in $S+S$ collisions, depends on the plasma lifetime. Our
calculations would attribute the observed enhancement to these
processes for a plasma lifetime of ca. 4~fm/$c$. Such a value is
supported by hydrodynamical calculations \cite{barz,Udo,shurehand}.
This places an upper limit on the plasma lifetime. For shorter times,
other mechanisms, including hadronization \cite{su3hadron} and final
state interactions must also be invoked.

Our paper is organized as follows: In Section~\ref{scatsec}, we first
give a brief review of essential formalism, then we discuss quark quark
scattering in the nonstrange and strange sectors. Following this, quark
antiquark scattering for both nonstrange and strange partons is
presented. Transport properties are derived and discussed in
Section~\ref{secthree}. The possibility of strangeness enhancement due
to quark antiquark scattering is discussed in Section~\ref{secfour}.
A general discussion of our methods and results is presented in
Section~\ref{sechen}. We summarize and conclude in
Section~\ref{secfive}.

\section{Scattering Processes} \label{scatsec}
\subsection{Review of Formalism} \label{formsec}
The model which we employ for our calculations is the $SU_f(3)$ version
of the Nambu--Jona--Lasinio (NJL) model\cite{sandi,hatkun} considered
at finite temperature. In this subsection, we give a brief review of the
basic formalism and define our notation, without giving derivations of
our formulae. For a more detailed treatment, the reader is referred to
Refs.~\cite{sandi,hatkun,su3hadron,loopies}. The Lagrangian for this
model is
\begin{eqnarray}
{\cal L} &=& \sum_{f=u,d,s} \bar\psi_f(i\partial\! \! \! /-m_{0f})\psi_f +
        G\sum_{a=0}^8\left[(\bar\psi\lambda^a\psi)^2 +
                          (\bar\psi i\gamma_5\lambda^a\psi)^2\right]
        \label{lagrange1} \\ \nonumber &-&
        K\left[\det\bar\psi(1+\gamma_5)\psi + \det\bar\psi(1-\gamma_5)\psi
                                \right] \quad,
\end{eqnarray}
where $G$ and $K$ are coupling constants with dimensions [MeV]$^{-2}$
and [MeV]$^{-5}$ respectively, and $m_{0f}$ are the current masses for
quarks of flavor $f$, which explicitly break chiral symmetry. The
determinantal term leads to the appearance of six fermion vertices,
which, in a mean field approximation, can be reduced to an effective
four point interaction by contracting out $\bar\psi\psi$
pairs\cite{sandi}. Under the additional assumption of $SU_f(2)$ isospin
symmetry, $m_u=m_d$, this leads to the effective Lagrangian
\begin{eqnarray}
{\cal L} &=& \sum_{f=u,d,s} \bar\psi_f(i\partial\! \! \! /-m_{0f})\psi_f
      + \sum_{a=0}^8\left[K_a^-(\bar\psi\lambda^a\psi)^2 +
                        K_a^+(\bar\psi i\gamma_5\lambda^a\psi)^2\right]
\label{lagrange2} \\ \nonumber &+&
        K_{08}^-\left[(\bar\psi\lambda^8\psi)(\bar\psi\lambda^0\psi)
              + (\bar\psi\lambda^0\psi)(\bar\psi\lambda^8\psi)\right]
        \nonumber \\ &+&
        K_{08}^+\left[(\bar\psi i\gamma_5\lambda^8\psi)
                (\bar\psi i\gamma_5\lambda^0\psi)
        +       (\bar\psi i\gamma_5\lambda^0\psi)
                (\bar\psi i\gamma_5\lambda^8\psi)\right] \nonumber
\end{eqnarray}
with the effective coupling constants
\begin{equation}
\begin{array}{ll}
K_0^\pm = G \mp \frac{1}{3} K (2{\cal G}^u+{\cal G}^s)
\quad, \qquad &
K_1^\pm=K_2^\pm=K_3^\pm = G\pm\frac{1}{2}K {\cal G}^s
\quad, \\
K_4^\pm=K_5^\pm = K_6^\pm=K_7^\pm  = G\pm\frac{1}{2}K {\cal G}^u
\quad, \qquad &
K_8^\pm = G\pm\frac{1}{6}K (4{\cal G}^u-{\cal G}^s)
\quad, \\
K_{08}^\pm= \pm \frac{\sqrt{2}}{6} K ({\cal G}^u-{\cal G}^s)
\quad, \qquad &
\end{array} \label{cpleff} \end{equation}
expressed in terms of the couplings $G$, $K$, and the trace of the
Green function $S_f(x,x)$ via
\begin{equation}
{\cal G}^f = N_c i{\rm tr}_\gamma S_f(x,x)
           = - \frac{N_c}{4\pi^2} m_f A(m_f, \mu_f) \quad,
\end{equation}
where, in the case of finite temperature and chemical potential,
\begin{equation}
A(m_f, \mu_f) =
\frac{16\pi^2}{\beta} \sum_n e^{i\omega_n\eta}
\int\limits_{|\vec p| < \Lambda} \frac{d^3p}{(2\pi)^3}
\frac{1}{(i\omega_n+\mu_f)^2-(\vec p ^2 + m_f^2)} \quad.
\label{Adef}
\end{equation}
In this expression, $\omega_n$ are the Matsubara frequencies for fermions,
$\omega_n=(2n+1)\pi/\beta$, and the sum runs over all positive and negative
integer values. Note that this quantity is derived from the finite temperature
Green function
\begin{equation}
S_f(\vec x - \vec x', \tau - \tau') = \frac{i}{\beta}\sum_n
e^{-i\omega_n(\tau-\tau')} \int \frac{d^3p}{(2\pi)^3}
\frac{e^{i\vec p(\vec x-\vec x')}}
{\gamma_0(i\omega_n+\mu_f) - \vec\gamma \vec p -m_f}
\quad.
\end{equation}
Explicit expressions for $A(m, \mu)$, which are suitable for numerical
evaluation, can be found in Ref.~\cite{loopies}. Note that we employ a
three momentum cutoff $\Lambda$ in order to make the integral in
Eq.~(\ref{Adef}) finite.
The physical quark masses $m_u$, $m_d$, $m_s$ are calculated from the
three coupled gap equations
\begin{equation}
\label{gapeq}
m_i = m_{0i} - \frac{GN_c}{\pi^2}m_iA(m_i, \mu_i)
             + \frac{KN_c^2}{8\pi^4}m_j A(m_j, \mu_j)m_kA(m_k, \mu_k)
\quad,
\end{equation}
where $i$, $j$, $k$ are three pairwise distinct flavors.
An explicit numerical solution of Eq.~(\ref{gapeq}) can be found in
Ref.~\cite{su3hadron}.

Mesons are constructed by calculating the quark--antiquark effective
interaction within the random phase approximation \cite{sandi,hatkun}.
In the pseudoscalar sector, this leads to the explicit form for the
pion and kaon propagators
\begin{equation}
{\cal D}_\pi(p_0, \vec p) = \frac{2K_3^+}{1-4K_3^+\Pi_{u\bar u}^P(p_0, \vec p)}
\label{piprop}
\end{equation}
\begin{equation}
{\cal D}_K(p_0, \vec p) = \frac{2K_4^+}{1-4K_4^+\Pi_{u\bar s}^P(p_0, \vec p)}
\quad. \label{kaprop}
\end{equation}
The irreducible pseudoscalar polarization function $\Pi_{f\bar f'}^P$
required in Eqs.~(\ref{piprop}) and (\ref{kaprop}) is given explicitly
by \cite{su3hadron}
\begin{eqnarray}
\Pi_{f\bar f'}^P(p_0, \vec p) &=& -\frac{N_c}{8\pi^2} \Big\{A(m_f,\mu_f) +
A(m_{f'}, \mu_{f'}) \label{polarihoe} \\
&+&\left[(m_f-m_{f'})^2-(p_0+\mu_f-\mu_{f'})^2 + \vec p^2\right]
B_0(\vec p, m_f, \mu_f, m_{f'}, \mu_{f'}, p_0)\Big\} \nonumber \quad,
\end{eqnarray}
where $B_0$ is the analytical continuation of
\begin{eqnarray}
& &B_0(\vec p, m_f, \mu_f, m_{f'}, \mu_{f'}, i\nu_m) =
\label{b0def} \\ & &
\frac{16\pi^2}{\beta} \sum_n e^{i\omega_n\eta}
\int\limits_{|\vec q| < \Lambda} \frac{d^3q}{(2\pi)^3}
\frac{1}{[(i\omega_n+\mu_f)^2-E_f^2]}
\frac{1}{[(i\omega_n-i\nu_m+\mu_{f'})^2-E_{f'}^2]}
\nonumber
\end{eqnarray}
($E_f=\sqrt{\vec q^2 + m_f^2}$, $E_{f'}=\sqrt{(\vec q - \vec p)^2 + m_{f'}^2}$)
to the real axis.
Explicit expressions for $B_0$ can again be found in \cite{loopies}.

For the $\eta$ and $\eta'$, the situation is somewhat more involved due to
the mixing terms in Eq.~(\ref{lagrange2}). The quark--antiquark
scattering matrix containing these particles is a $2\times 2$ matrix
\begin{equation}
{\cal D}_{\eta, \eta'} = 2\frac{\cal K}{{\cal AC-B}^2}
                         \left( \begin{array}{cc} {\cal A} & {\cal B} \\
                                {\cal B} & {\cal C} \end{array} \right)
\label{cocktail}
\end{equation}
with
\begin{mathletters}
\label{etaprop}
\begin{equation}
{\cal A} = K_0^+ - \frac{4}{3} {\cal K} (\Pi^P_{u\bar u}+2\Pi^P_{s\bar s})
\end{equation} \begin{equation}
{\cal B} = K_{08}^+ +
\frac{4\sqrt{2}}{3}{\cal K}(\Pi^P_{u\bar u}-\Pi^P_{s\bar s})
\end{equation} \begin{equation}
{\cal C} = K_8^+ - \frac{4}{3} {\cal K} (2\Pi^P_{u\bar u}+\Pi^P_{s\bar s})
\end{equation} \begin{equation}
{\cal K} = K_0^+K_8^+-K_{08}^2 \quad.
\end{equation}
\end{mathletters}
Since we do not require the meson masses and their static properties
explicitly in this work,
we do not display the technical details needed for this here, but rather
refer our reader to Ref.~\cite{su3hadron}.

In the scalar sector, the NJL model in its $SU_f(3)$ version contains
nine mesons ($\sigma_{\pi^0}$, $\sigma_{\pi^\pm}$, $\sigma_{K^0}$,
$\sigma_{\overline{K^0}}$, $\sigma_{K^\pm}$, $\sigma$, $\sigma'$), that
accompany the nine mesons ($\pi^0$, $\pi^\pm$, $K^0$, $\overline{K^0}$,
$K^\pm$, $\eta$, $\eta'$) in the pseudoscalar sector.  The propagators
for the scalar particles can be immediately obtained from
Eqs.~(\ref{piprop})--(\ref{etaprop}) by replacing the pseudoscalar
coupling strengths $K_i^+$ by the scalar ones $K_i^-$ and replacing the
pseudoscalar polarization by the scalar polarization
\begin{eqnarray}
\Pi_{f\bar f'}^S(p_0, \vec p) &=& -\frac{N_c}{8\pi^2} \Big\{A(m_f,\mu_f) +
A(m_{f'}, \mu_{f'}) \label{scalprop} \\
&+&\left[(m_f+m_{f'})^2-(p_0+\mu_f-\mu_{f'})^2 + \vec p^2\right]
B_0(\vec p, m_f, \mu_f, m_{f'}, \mu_{f'}, p_0)\Big\} \nonumber \quad.
\end{eqnarray}
Via Eqs.~(\ref{piprop}), (\ref{kaprop}) and (\ref{cocktail}), the
masses of the pseudoscalar mesons as well as those of the scalar mesons
with the appropriate changes mentioned above, have been obtained as a
function of temperature. For explicit values see Ref.~\cite{su3hadron}.

\subsection{Quark Quark Scattering}
In this subsection, we classify and illustrate via calculation the
possible independent elastic scattering collision processes within the
$SU_f(3)$ flavor quark--quark combinations.

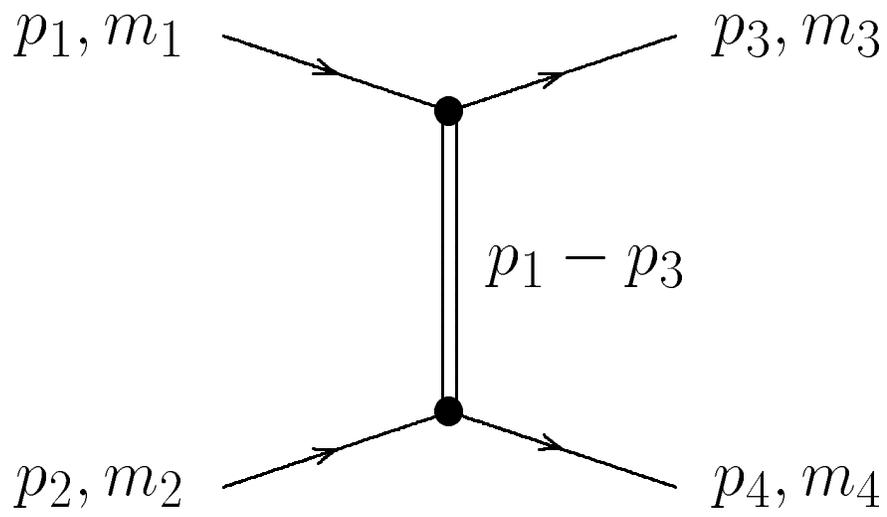
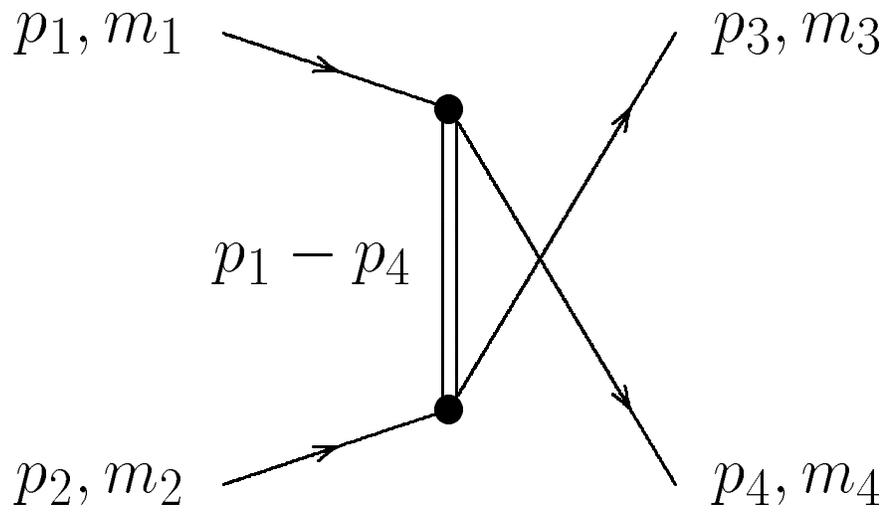
\begin{figure}
\caption[]{Feynman diagrams for elastic $qq$ scattering to lowest order
        in $1/N_c$.}
\label{qqscat}
\end{figure}

For elastic scattering processes of the type $qq'\to qq'$, via simple
combinatorics we have six different possibilities available. However,
due to isospin symmetry, the processes $us\to us$ and $ds\to ds$ lead
to the same cross section.  The same holds true for the processes
$uu\to uu$ and $dd\to dd$, so that we have in total only four
independent processes for quark--quark scattering. These are listed in
Table~\ref{qqtab}. The Feynman diagrams appropriate for such scattering
in lowest order in terms of a $1/N_c$ expansion, where $N_c$ is the
number of colors, are shown in Fig.~\ref{qqscat}. Here both
$t$--channel and $u$--channel diagrams can occur, with the species of
the exchanged meson depending on the specific process in question.
These are also given explicitly in Table~\ref{qqtab} for each of the
independent processes.  We note that there is {\em at least\/} one
scalar and one pseudoscalar exchange channel for each diagram. In this
point, this $SU_f(3)$ chiral model differs markedly from the $SU_f(2)$
chiral model, as is evident from the fact that the $SU_f(2)$ model
supports only one scalar meson. This will be seen explicitly in the
numerical calculations.

In the following, we will derive as far as possible general expressions
for the cross sections. We illustrate these with explicit calculations
for the specific process $uu\to uu$.

\subsubsection{Analytical Calculations}
The matrix elements corresponding to the Feynman graphs of
Fig.~\ref{qqscat} can in general be written as
\begin{eqnarray}
-i{\cal M}_t &=& \delta_{c_1,c_3} \delta_{c_2,c_4}
                 \bar u(p_3) T u(p_1)
                 \left[i{\cal D}_t^S(p_1-p_3)\right]
                 \bar u(p_4) T u(p_2)
\nonumber \\
             &+& \delta_{c_1,c_3} \delta_{c_2,c_4}
                 \bar u(p_3)(i\gamma_5 T)u(p_1)
                 \left[i{\cal D}_t^P(p_1-p_3)\right]
                 \bar u(p_4)(i\gamma_5 T)u(p_2)
\label{tchan}
\end{eqnarray}
\begin{eqnarray}
-i{\cal M}_u &=& \delta_{c_1,c_4} \delta_{c_2,c_3}
                 \bar u(p_4) T u(p_1)
                 \left[i{\cal D}_u^S(p_1-p_4)\right]
                 \bar u(p_3) T u(p_2)
\nonumber \\
             &+& \delta_{c_1,c_4} \delta_{c_2,c_3}
                 \bar u(p_4)(i\gamma_5 T)u(p_1)
                 \left[i{\cal D}_u^P(p_1-p_4)\right]
                 \bar u(p_3)(i\gamma_5 T)u(p_2)
\quad. \label{uchan}
\end{eqnarray}
Here $T {\cal D}_t^S T$ is a symbolic expression for the
sum over all scalar $t$ exchange channels of flavor factors times
particle propagator, with $T {\cal D}_t^P T$ being the same for
pseudoscalar $t$ graphs.  The symbols $T {\cal D}_u^S T$
and $T {\cal D}_u^P T$ are the analogous quantities for the
$u$--channel. Note that the matrix elements (\ref{tchan}) and (\ref{uchan})
describe the scattering due to the exchange of colorless mesons.
The scattering cross sections due to the exchange of color octet states
can be estimated to differ from Eqs.~(\ref{tchan}), (\ref{uchan}) by a
factor 4/9 \cite{su2elast} and are neglected here.

To calculate the cross sections, we require the squared matrix
elements, summed over spin and color degrees of freedom and averaged
over the incoming states,
\begin{equation}
\frac{1}{4N_c^2} \sum_{s, c} \left| {\cal M}_t
        - {\cal M}_u \right|^2 \quad.
\end{equation}
General expressions for these, apart from the flavor factors stemming
from the Gell--Mann matrices, are given in Appendix~\ref{ampliapp}. The
flavor factors, together with the propagators of the exchanged mesons,
have to be specified for each individual process. We illustrate this
procedure explicitly here for the process $uu\to uu$. From
Table~\ref{qqtab}, we read off that the $t$--channel exchange proceeds
via an $\pi^0$, $\eta$ and $\eta'$ in the pseudoscalar sector and via
a $\sigma_{\pi^0}$, $\sigma$
and $\sigma'$ in the scalar sector.  This means that the propagators
for the $t$--channel have the form
\begin{equation}
T {\cal D}^P_t T = {\cal D}_\pi + \frac{4}{3} \frac{\cal K}{{\cal AC-B}^2}
\left({\cal A}+\sqrt{2}{\cal B} + \frac{1}{2} {\cal C} \right) \quad,
\label{uuprop}
\end{equation}
where ${\cal D}_\pi$ is defined in Eq.~(\ref{piprop}) and
${\cal A}$, ${\cal B}$, ${\cal C}$ and ${\cal K}$ are given by
Eq.~(\ref{etaprop}). The propagator for the scalar $t$--channel
has the same form with, however, the coupling constants and polarization
functions suitably altered, as was discussed in Sec.~\ref{formsec}.
The $u$--channel exchange proceeds via the exchange of the same
particles. Thus for ${\cal D}_u$ we have the same form as in
Eq.~(\ref{uuprop}).

In calculating the cross section, we confine ourselves to the situation
in which the center of mass system of the incoming particles is at rest
relative to the medium. In addition, in this paper, we will examine the
cross sections at finite temperature, and in what follows, set the
chemical potential to zero.  In this case the differential cross
section can be written as
\begin{equation}
\frac{d\sigma}{dt} = \frac{1}{16 \pi s (s-4m_u^2)}
\frac{1}{4N_c^2} \sum_{s, c} \left| {\cal M}_t
        - {\cal M}_u \right|^2 \quad.
\end{equation}
For the total cross section, we include a Fermi blocking factor for
the final states
\begin{equation}
\sigma = \int dt \frac{d\sigma}{dt} [1-f_F(\beta \sqrt{s}/2)]^2
\quad,
\end{equation}
with the Fermi function $f_F(x)=1/(\exp x + 1)$. Note that the total
cross section depends only on the temperature $T$ and the center of
mass energy $\sqrt{s}$ \cite{su2elast}.

The process $uu\to uu$ is an example of the elastic scattering of
quarks that is already present in the $SU_f(2)$ model. It differs from
this three flavor calculation in that the additional exchange channels
$\sigma_\pi$, $\eta$, $\eta'$ and $\sigma'$ are missing in that case.
For a pure $SU_f(2)$ calculation, only the terms
\begin{mathletters} \begin{eqnarray}
T {\cal D}_t^P T &=& \frac{2G}{1-4G\Pi_{u\bar u}^P(0, \sqrt{-t})} \\
T {\cal D}_t^S T &=& \frac{2G}{1-4G\Pi_{u\bar u}^S(0, \sqrt{-t})} \\
T {\cal D}_u^P T &=& \frac{2G}{1-4G\Pi_{u\bar u}^P(0, \sqrt{-u})} \\
T {\cal D}_u^S T &=& \frac{2G}{1-4G\Pi_{u\bar u}^S(0, \sqrt{-u})}
\end{eqnarray} \end{mathletters}
occur, leading us to the result of Ref.~\cite{su2elast}. The appearance
of the new exchange channels in both pseudoscalar and scalar sectors
leads to a significant enhancement in the numerical values of the cross
sections, as will be shown below.

One notes from Table~\ref{qqtab} that the $ud\to ud$ cross section
follows via exchange of the same scalar and pseudoscalar mesons in the
$t$--channel, but a restricted set in the $u$--channel is admitted. The
elastic $ss\to ss$ processes couple again to the same set of scalar and
pseudoscalar mesons in both $t$-- and $u$--channels. In this case,
no coupling to the $\pi$ or $\sigma_\pi$ can occur. When one light and
one heavy quark are scattered, one finds that virtual $K$ and
$\sigma_K$ exchanges are possible in the $u$--channel. This leads to
some technical difficulties that are due to the kinematics.
One would expect that the kaon (and $\sigma_K$) propagator
should not depend on the sign of the zero component of the exchanged
momentum. This corresponds, via Eqs.~(\ref{piprop}), (\ref{kaprop}), to
the condition $\Pi_{q\bar s}(p_0, \vec p) = \Pi_{q\bar s}(-p_0, \vec
p)$, or, through Eqs.~(\ref{polarihoe}), (\ref{scalprop}), to the
equivalent statement
\begin{equation}
B_0(\vec p, m_q, \mu_q, m_s, \mu_s, p_0) = B_0(\vec p, m_q, \mu_q,
m_s, \mu_s, - p_0) \quad. \label{symcond}
\end{equation}
However, this symmetry is lost in a straightforward evaluation
\cite{loopies}, which is an artifact of our regularization procedure.
This can be explicitly traced back to the fact that a shift in variable
is performed in certain terms occurring in $\Pi_{q\bar s}$. Such
variable shifts are strictly non admissable for divergent integrals,
but since they contribute only to order $1/\Lambda$, they are usually
ignored.  The symmetry condition (\ref{symcond}) is in any event still
fulfilled if the constituent quark masses are equal (i.e. for pions,
etas and their scalar partners) or the three momentum argument
vanishes.  For evaluations in the kaonic sector, however, the violation
of (\ref{symcond}) becomes evident. Accompanying this artifact is the
further cutoff and regularization scheme induced artifact that the
imaginary part of the polarization is discontinous at $p_0=0$, i.e. at
the kinematical point that is required e.g. for the process $u\bar u\to
s\bar s$ in the center of mass system. We resolve this problem by hand
by replacing $B_0(\vec p, m_q, \mu_q, m_s, \mu_s, p_0)$ with the
symmetric form
\begin{equation}
\frac{1}{2} \left[ B_0(\vec p, m_q, \mu_q, m_s, \mu_s, p_0) +
B_0(\vec p, m_q, \mu_q, m_s, \mu_s, -p_0) \right]
\end{equation}
whenever dealing with kaons.

\subsubsection{Numerical Results}
Our numerical calculations were performed using the parameter set
$m_{0u}=m_{0d}=5.5~{\rm MeV}$, $m_{0s}=140.7~{\rm MeV}$,
$G\Lambda^2=1.835$, $K\Lambda^5=12.36$ and $\Lambda=602.3~{\rm MeV}$.
This is the same parameter set that was used in Ref.~\cite{su3hadron},
so the numerical results concerning the static mesonic properties
can be obtained from this reference. In particular it was demonstrated in
Ref.~\cite{su3hadron} that at the pionic Mott temperature
$T_{M\pi}=212~{\rm MeV}$, the pion mass becomes equal to the mass
of its constituents $m_\pi=2m_u$ and the pion becomes a resonant
state. The same happens with the kaon at $T_{MK}=210~{\rm MeV}$,
the $\eta$ at $T_{M\eta}=180~{\rm MeV}$, and the $\sigma$ at
$T_{M\sigma}=165~{\rm MeV}$. At these temperatures, the respective
particles become unbound. This effect in a rather crude fashion models
the deconfinement transition within the NJL model \cite{gerry}.

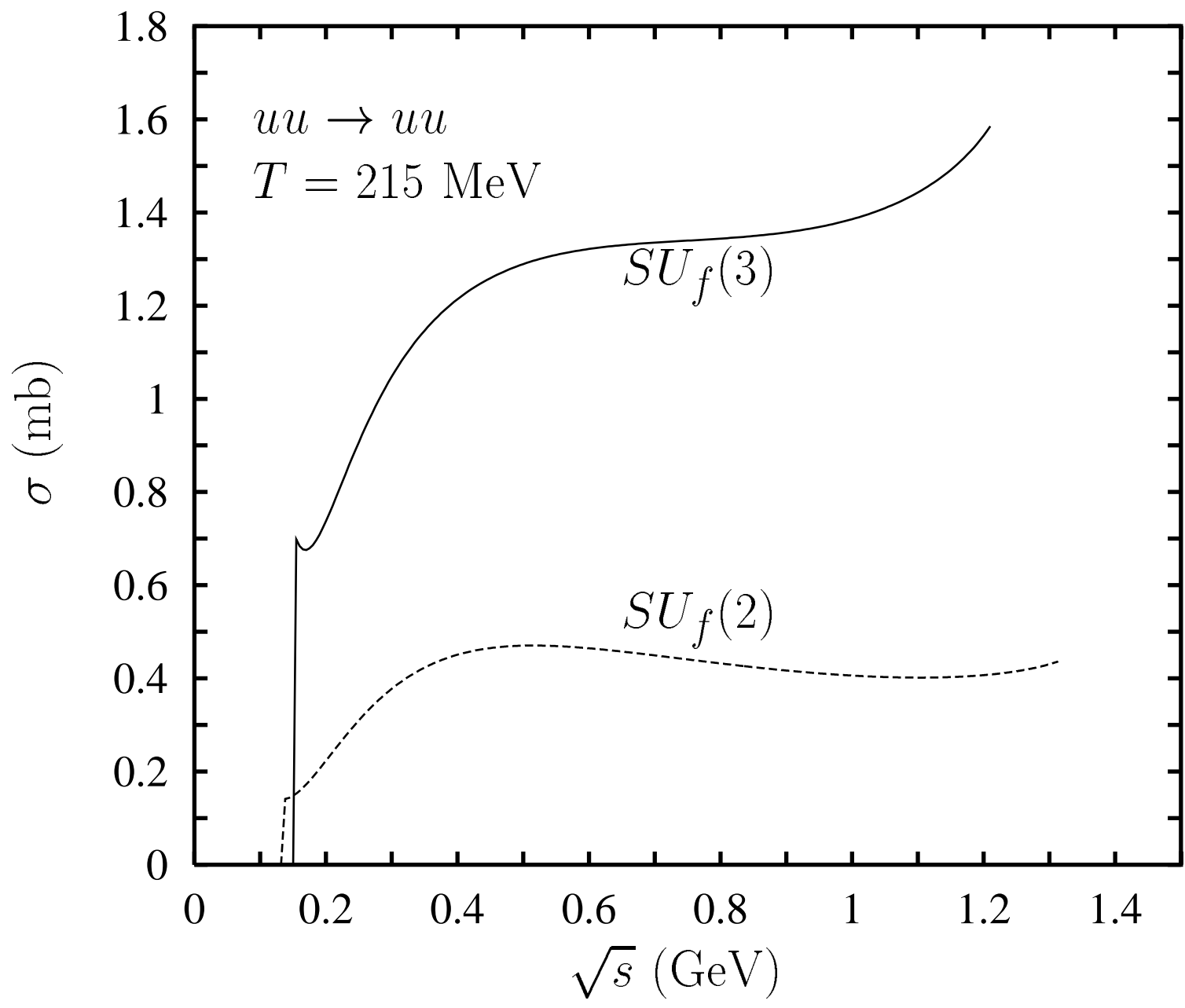
\begin{figure}
\caption[]{Comparision of the cross section for $uu\to uu$
	for three flavors (solid line) with the $SU_f(2)$ result of
	Ref.~\cite{su2elast} (dashed line), at $T=215~{\rm MeV}$.}
\label{su23vergl}
\end{figure}

We first compare our results for three flavors with the corresponding
calculation of Ref.~\cite{su2elast}. In Fig.~\ref{su23vergl}, we show
the total cross section for the process $uu\to uu$ at $T=215~{\rm MeV}$
as a function of $\sqrt{s}$. The temperature chosen lies slightly higher
than the pion Mott temperature $T_{M\pi}$; we are therefore in the
plasma phase, in which the model may be regarded as physically
realistic. As in Ref.~\cite{su2elast}, our calculations are shown for
center of mass energies $\sqrt{s}\le 2\sqrt{\Lambda^2+m_u^2}$, i.e. they
are restricted by the natural cutoff of this model. Numerically we find
that the $SU_f(3)$ calculation yields a far larger cross section for
$uu\to uu$ than the corresponding two flavor case, the difference being
a factor of 3--4. Thus the greater number of exchange channels that
become available for three flavors of quarks significantly enhances
these scattering processes, and the $SU_f(2)$ result is only recovered
on directly eliminating these. The remainder of our calculations are
shown for the three flavor sector only.

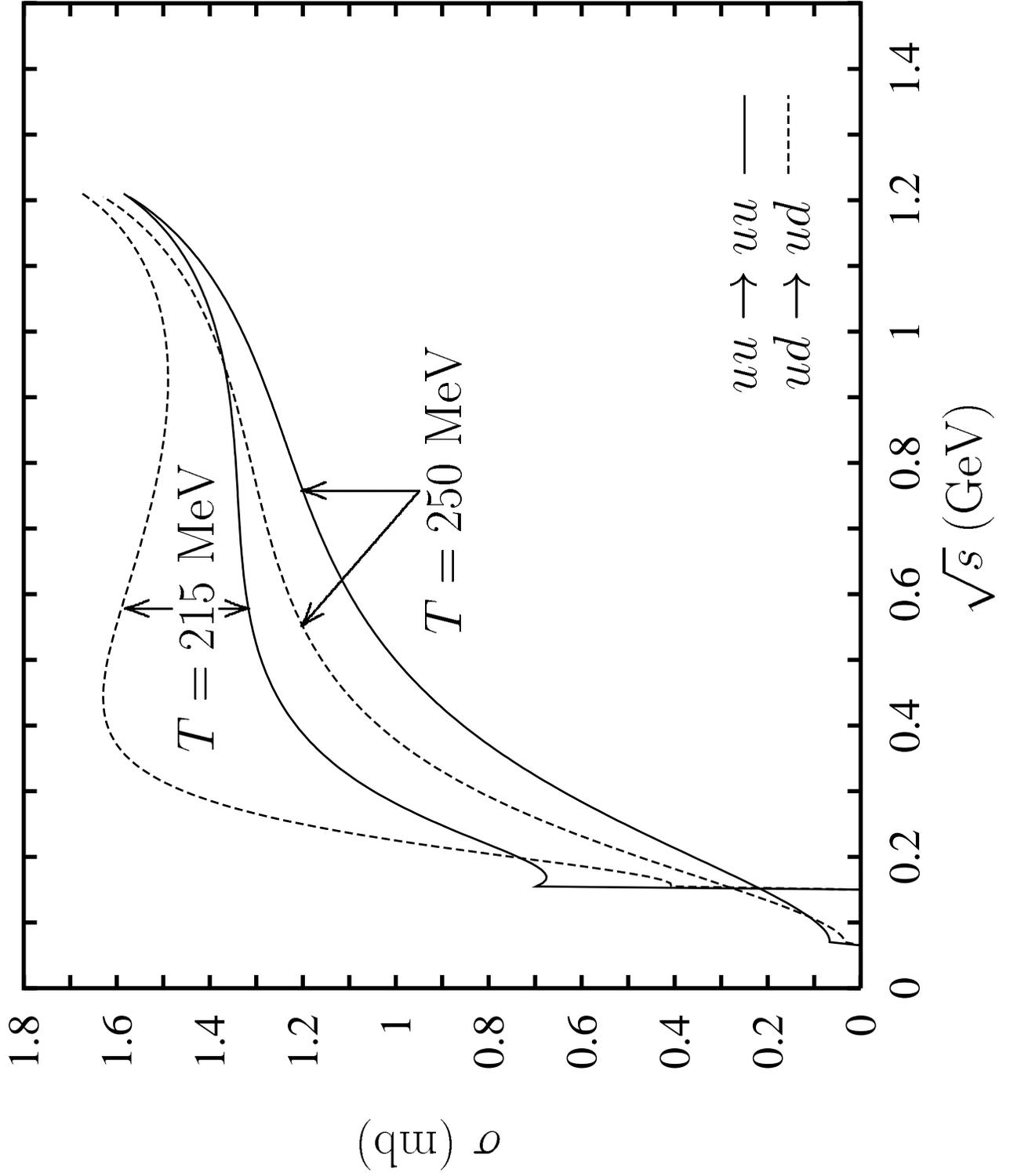
\begin{figure}
\caption[]{Total cross section for elastic $qq$ scattering involving
	only light quarks, as a function of $\sqrt{s}$, at $T=215{\rm
	MeV}$ and $T=250{\rm MeV}$.}
\label{qqlite}
\end{figure}

In Fig.~\ref{qqlite}, we show the total $qq$ scattering cross sections
that involve nonstrange quarks as a function of $\sqrt{s}$ and for two
values of the temperature, $T=215~\mbox{MeV}$ and $T=250~\mbox{MeV}$.
The curves for both $uu\to uu$ and $ud\to ud$ are essentially flat and
display no particular structures. The $ud\to ud$ scattering cross
section lies at a given temperature slightly higher than the $uu\to
uu$ cross section, with both between 1.2 and 1.6~mb for
$T=215~\mbox{MeV}$.  This difference can be attributed to the different
flavor factors accompanying the various mesonic states. Note that the
$ud$ scattering in fact has less exchange mesons available in the
$u$--channel, since no neutral particles are admissible in this
channel. As the temperature is increased from $T=215~\mbox{MeV}$ to
$T=250~\mbox{MeV}$, the cross sections become somewhat smaller in
magnitude.  One also sees that the thresholds for quark quark
scattering have shifted to slightly lower values, due to the
temperature dependence of the quark masses themselves.

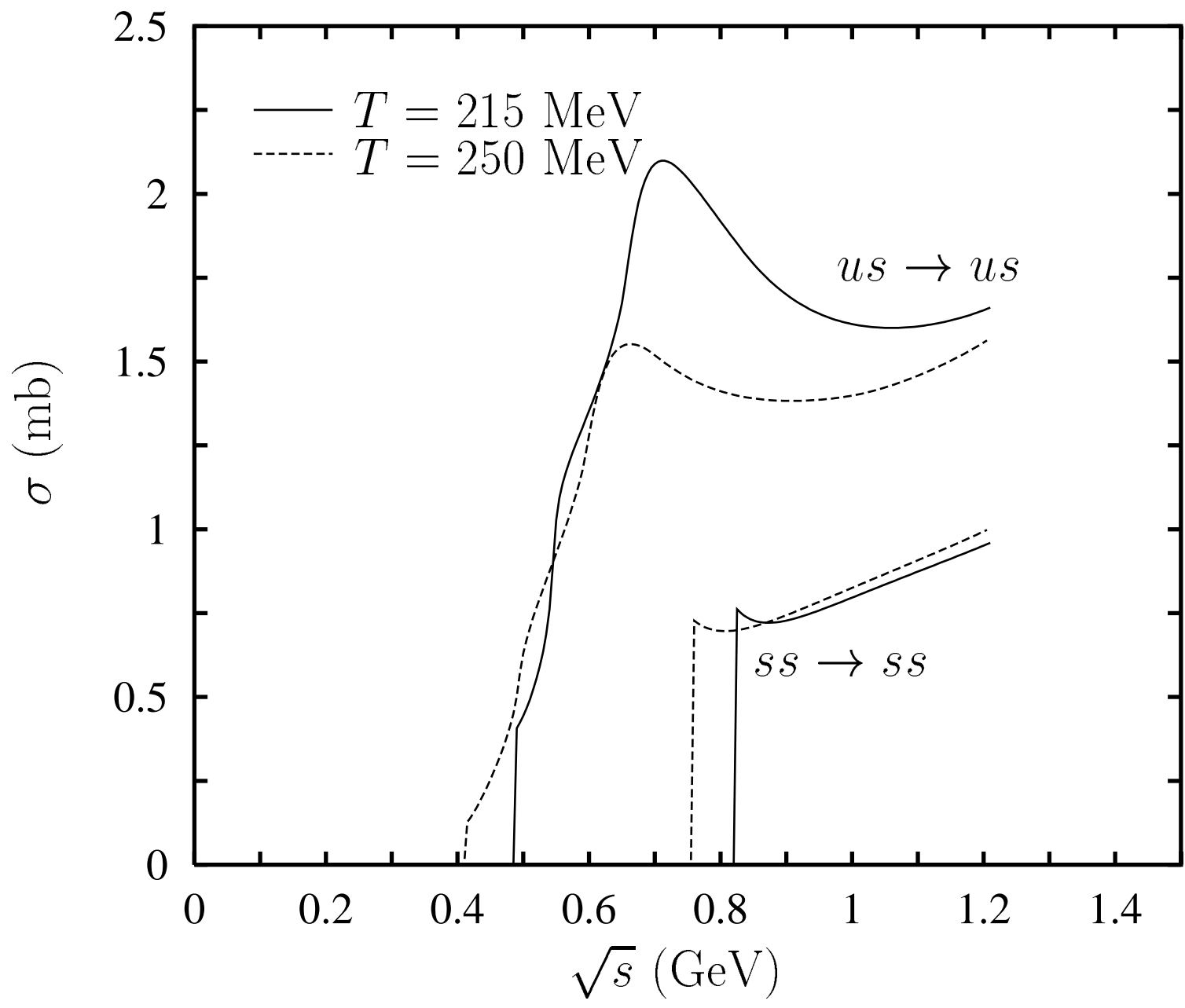
\begin{figure}
\caption[]{Total cross section for elastic $qq$ scattering involving at
	least one strange quark, as a function of $\sqrt{s}$, at
	$T=215{\rm MeV}$ (solid lines) and $T=250{\rm MeV}$ (dashed
	lines).}
\label{qqstra}
\end{figure}

Quark quark scattering processes that involve at least one strange
quark are shown in Fig.~\ref{qqstra} as a function of $\sqrt{s}$ for
the temperatures $T=215~\mbox{MeV}$ and $T=250~\mbox{MeV}$. The peak
like structure in the $us\to us$ scattering cross section is a cutoff
artifact. The cross sections are otherwise featureless. Increasing the
temperature from $T=215~\mbox{MeV}$ to $T=250~\rm{MeV}$ reduces the
$us\to us$ cross section slightly, while raising the $ss\to ss$ by a
small amount.

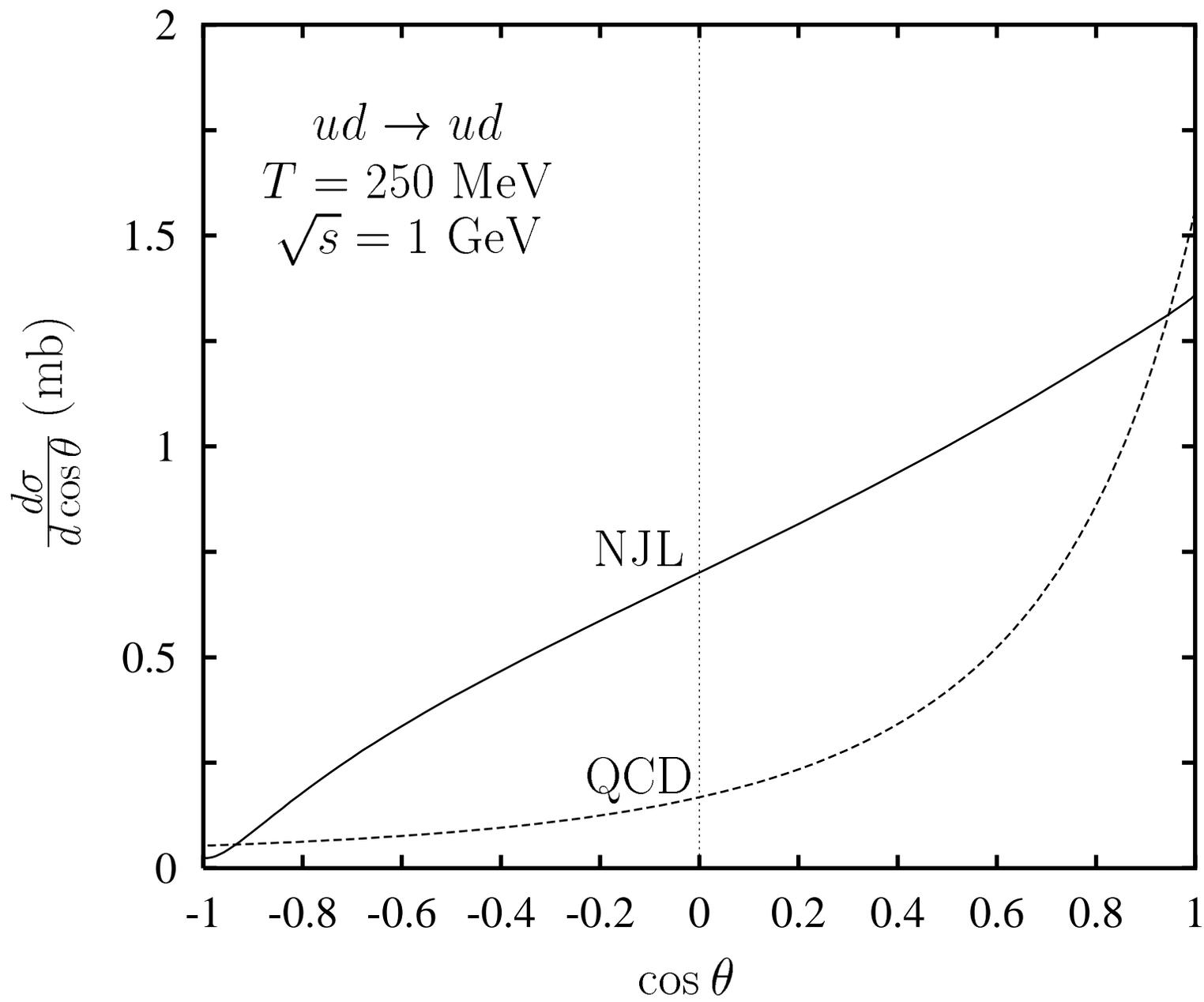
\begin{figure}
\caption[]{Differential cross section for $ud\to ud$ at $T=250{\rm MeV}$
        and $\sqrt{s}=1~{\rm GeV}$ within the NJL model (solid line)
        and perturbative QCD (dashed line).}
\label{diffqq}
\end{figure}

Differential cross sections may also be calculated for each process. As
an example, we show the differential cross section
$d\sigma/d\cos\theta$ for the process $ud\to ud$ as a function of
$\cos\theta$ at $\sqrt{s}=1~\mbox{GeV}$ and temperature
$T=250~\mbox{MeV}$ in Fig.~\ref{diffqq}. For comparison, we also
include this quantity calculated to lowest order in perturbative QCD
and using finite values of the quark masses as are indicated by the NJL
model at this temperature. An explicit formulation of the perturbative
QCD cross section with finite masses is given in Appendix~\ref{appqcd}.
Note that, since these cross sections are infrared divergent at this
level, we have introduced an effective gluon mass \cite{eddie}
\begin{equation}
m_g^2 = 2\pi\alpha_s\left(1+\frac{N_f}{6}\right) T^2 \approx
(600~{\rm MeV})^2 \quad,
\end{equation}
as a regulator, with $N_f=3$ being the number of flavors here and the
QCD coupling strength $\alpha_s$ taken {\em ad hoc\/} to be
$\alpha_s=0.6$. The perturbative QCD cross section, which has
essentially the same form as the M{\o}ller scattering cross section for
$e^+e^-$ scattering, displays a strong forward peak. In the limit
$m_g\to 0$, this peak becomes a pole and gives rise to the Coulomb
singularity well known from QED. The NJL cross section, on the other
hand, also displays a preference for the forward direction. However,
this maximum is not as pronounced.

\subsection{Quark Antiquark Scattering}
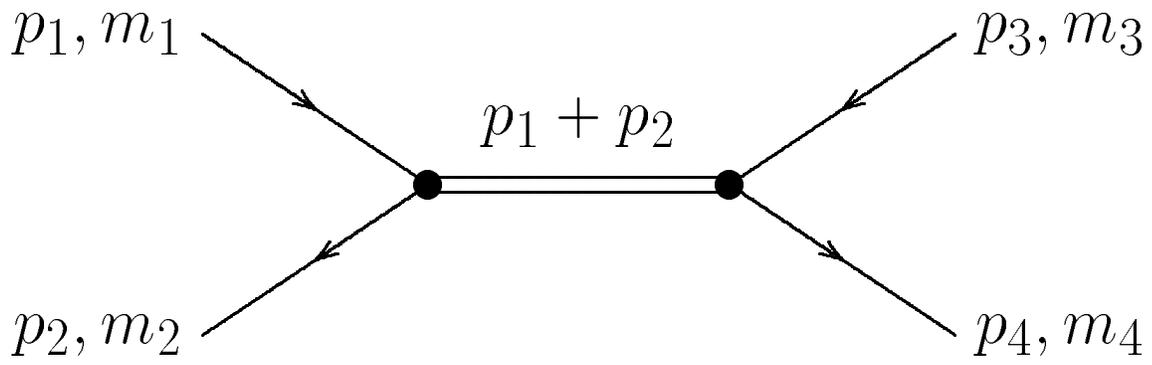
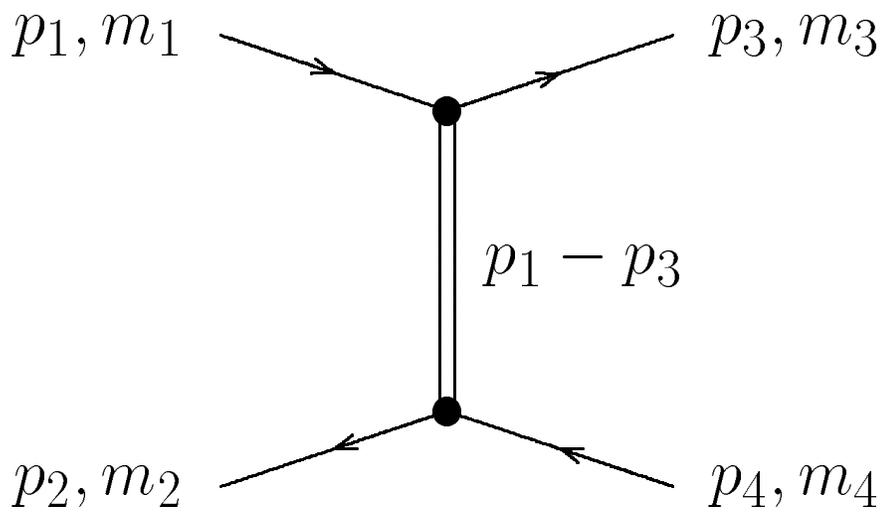
\begin{figure}
\caption[]{Feynman diagrams for elastic $q\bar q$ scattering to lowest
        order in $1/N_c$.}
\label{qqbscat}
\end{figure}

We now turn to a discussion of processes of the form $q\bar q'\to q\bar
q'$.  For these, one has seven independent processes out of a total
possible number of fifteen, taking again isospin and charge conjugation
symmetry into account.  These processes are listed in
Table~\ref{qqbtab}. The number of independent processes could be
further reduced by taking into account crossing symmetry, e.g. by
regarding $u\bar s\to u\bar s$ and $u\bar u\to s\bar s$ as dependent
processes. In this paper, however, we regard processes, which are
related by crossing symmetry as independent, since crossing symmetry
does not lead to numerically equal cross sections.  The relevant
Feynman diagrams are shown in Fig.~\ref{qqbscat}. Here only
$s$--channel and $t$--channel diagrams occur. The species of the
exchanged mesons are listed in Table~\ref{qqbtab} for the independent
processes.

As before, we will derive general expressions for the cross sections,
illustrating these with the example $u\bar s\to u\bar s$.

\subsubsection{Analytical Calculations}
The Feynman diagrams for $q\bar q$ scattering are shown in Fig.~\ref{qqbscat}.
Analogously to Eqs.~(\ref{tchan}), (\ref{uchan}), the transition amplitude
is given by
\begin{eqnarray}
-i{\cal M}_s &=& \delta_{c_1,c_2} \delta_{c_3,c_4}
                 \bar v(p_2) T u(p_1)
                 \left[i{\cal D}_s^S(p_1+p_2)\right]
                 \bar u(p_3) T v(p_4)
\nonumber \\
             &+& \delta_{c_1,c_2} \delta_{c_3,c_4}
                 \bar v(p_2)(i\gamma_5 T)u(p_1)
                 \left[i{\cal D}_s^P(p_1+p_2)\right]
                 \bar u(p_3)(i\gamma_5 T)v(p_4)
\end{eqnarray} \begin{eqnarray}
-i{\cal M}_t &=& \delta_{c_1,c_3} \delta_{c_2,c_4}
                 \bar u(p_3) T u(p_1)
                 \left[i{\cal D}_t^S(p_1-p_3)\right]
                 \bar v(p_2) T v(p_4)
\nonumber \\
             &+& \delta_{c_1,c_3} \delta_{c_2,c_4}
                 \bar u(p_3)(i\gamma_5 T)u(p_1)
                 \left[i{\cal D}_t^P(p_1-p_3)\right]
                 \bar v(p_2)(i\gamma_5 T)v(p_4) \quad.
\end{eqnarray}
The squares of these amplitudes can again be found in Appendix~\ref{ampliapp}.

We illustrate the further calculation by choosing the specific process
$u\bar s\to u\bar s$. As can be seen from Table~\ref{qqbtab}, the $s$
channel of this process proceeds via the exchange of a kaon in the
pseudoscalar and a $\sigma_K$ in the scalar part. This gives
\begin{eqnarray}
T {\cal D}_s^P T &=& 2 {\cal D}_K \\
T {\cal D}_s^S T &=& 2 {\cal D}_{\sigma_K} \quad.
\end{eqnarray}
The $t$ channel proceeds via the exchange of an $\eta$ or $\eta'$
in the pseudoscalar, a $\sigma$ or $\sigma'$ in the
scalar part. This means that we have
\begin{equation}
T {\cal D}_t^P T = \frac{4}{3} \frac{\cal K}{{\cal AC-B}^2}
\left({\cal A} - \frac{1}{\sqrt{2}}{\cal B} - {\cal C} \right)
\end{equation}
and an analogous expression for the scalar part.
The differential cross section is now
\begin{equation}
\frac{d \sigma}{dt}=\frac{1}{16\pi [s-(m_u+m_s)^2] [s-(m_u-m_s)^2]}
\frac{1}{4N_c^2} \sum_{s,c} \left|{\cal M}_s - {\cal M}_t \right|^2
\end{equation}
and the total cross section is calculated as
\begin{equation}
\sigma = \int dt \frac{d \sigma}{dt}
\left[1-f_F\left(\beta E_3\right)\right]
\left[1-f_F\left(\beta E_4\right)\right]
\quad, \label{cross}
\end{equation}
where a Fermi blocking factor for the final states has been introduced.
In Eq.~(\ref{cross}), $E_i^2=p_i^2+m_i^2$, where $i=3,4$.

\subsubsection{Numerical Results}
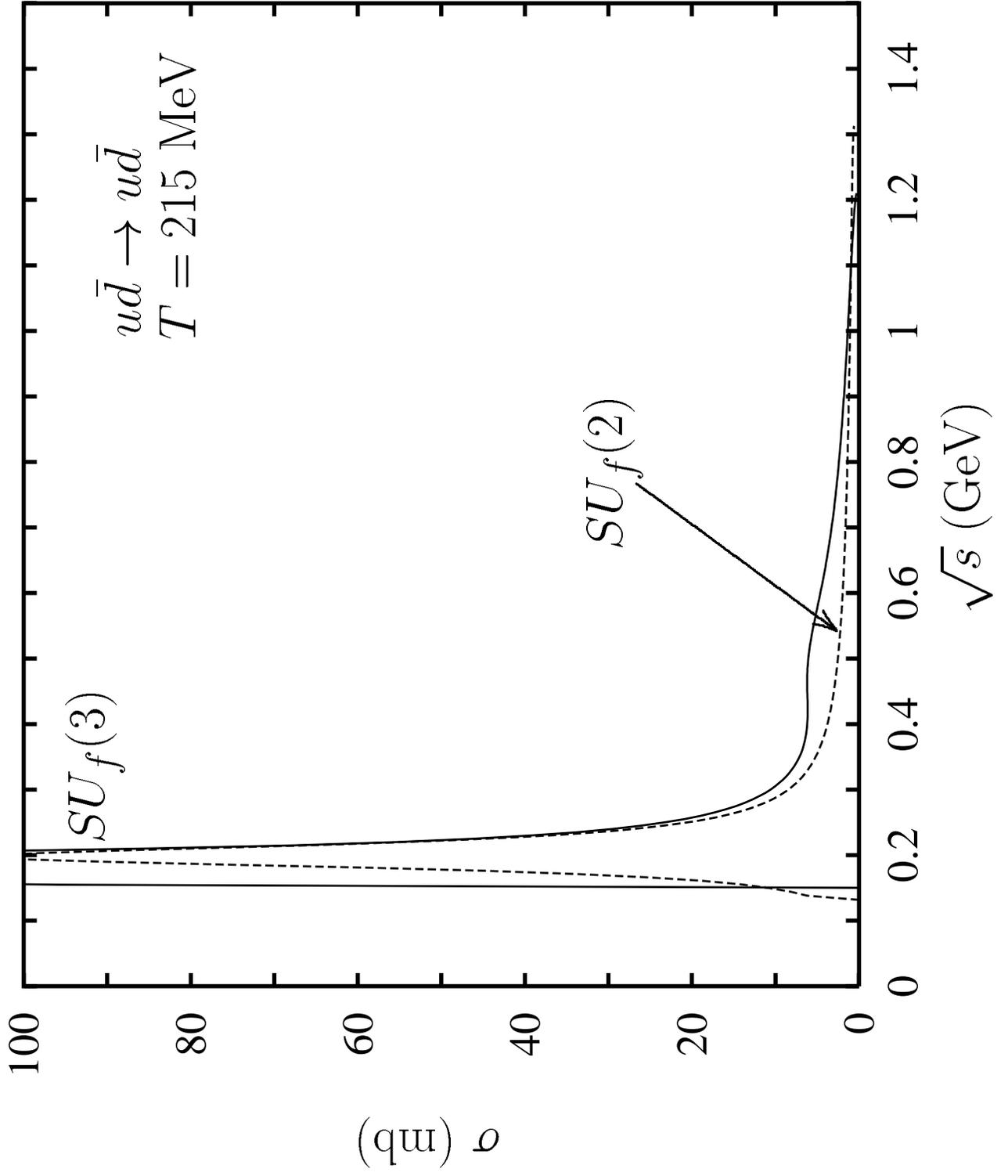
\begin{figure}
\caption[]{Comparison of the cross section $u\bar d\to u\bar d$
	calculated as a function of $\sqrt{s}$ and at
	$T=215~\mbox{MeV}$ for both $SU_f(3)$ (solid line) and
	$SU_f(2)$ (dashed line).}
\label{qqb23vergl}
\end{figure}

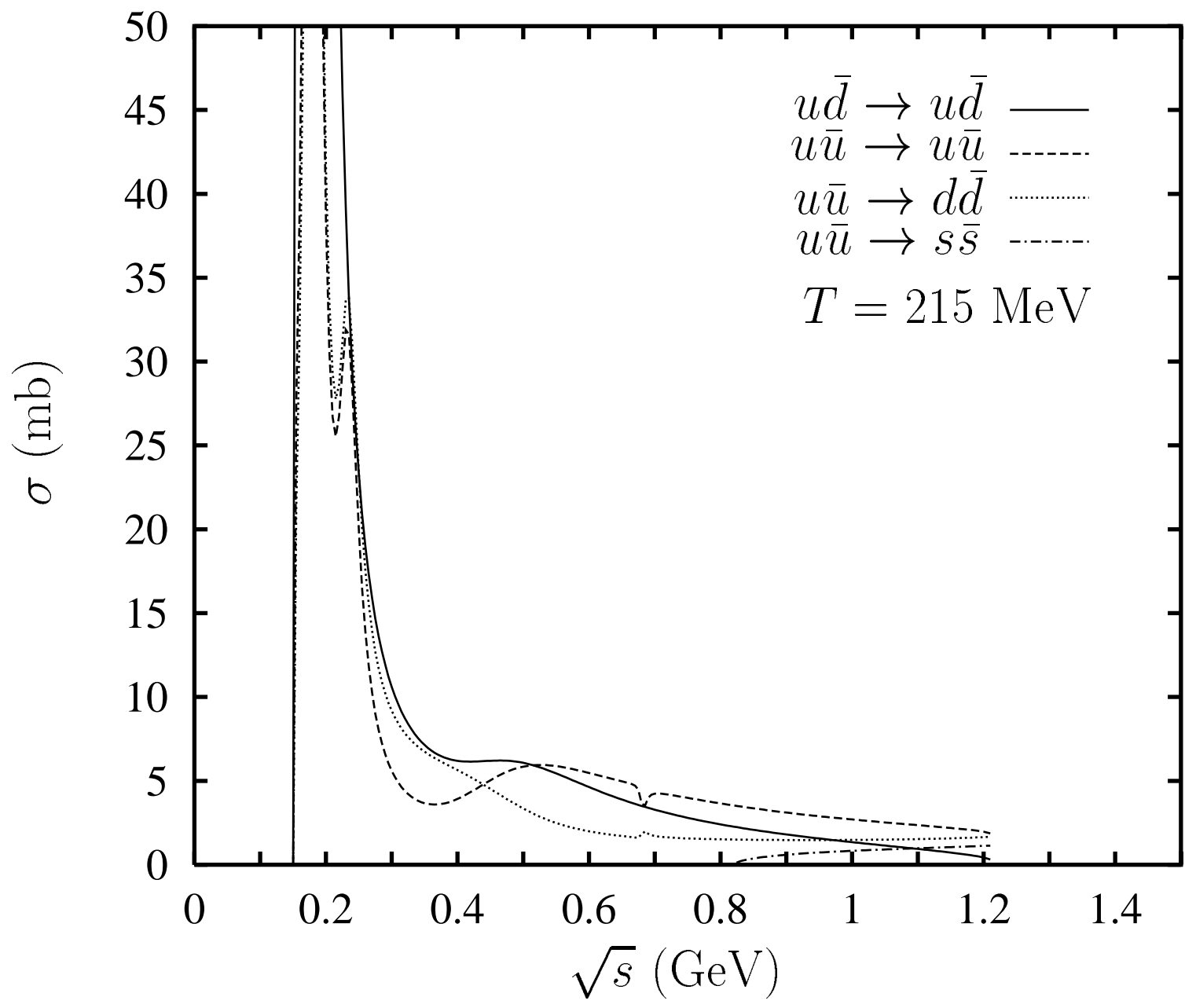
\begin{figure}
\caption[]{Total cross section for $q\bar q$ scattering with only light quarks
	in the initial state, shown as a function of $\sqrt{s}$, at
	$T=215{\rm MeV}$.}
\label{qqblite}
\end{figure}

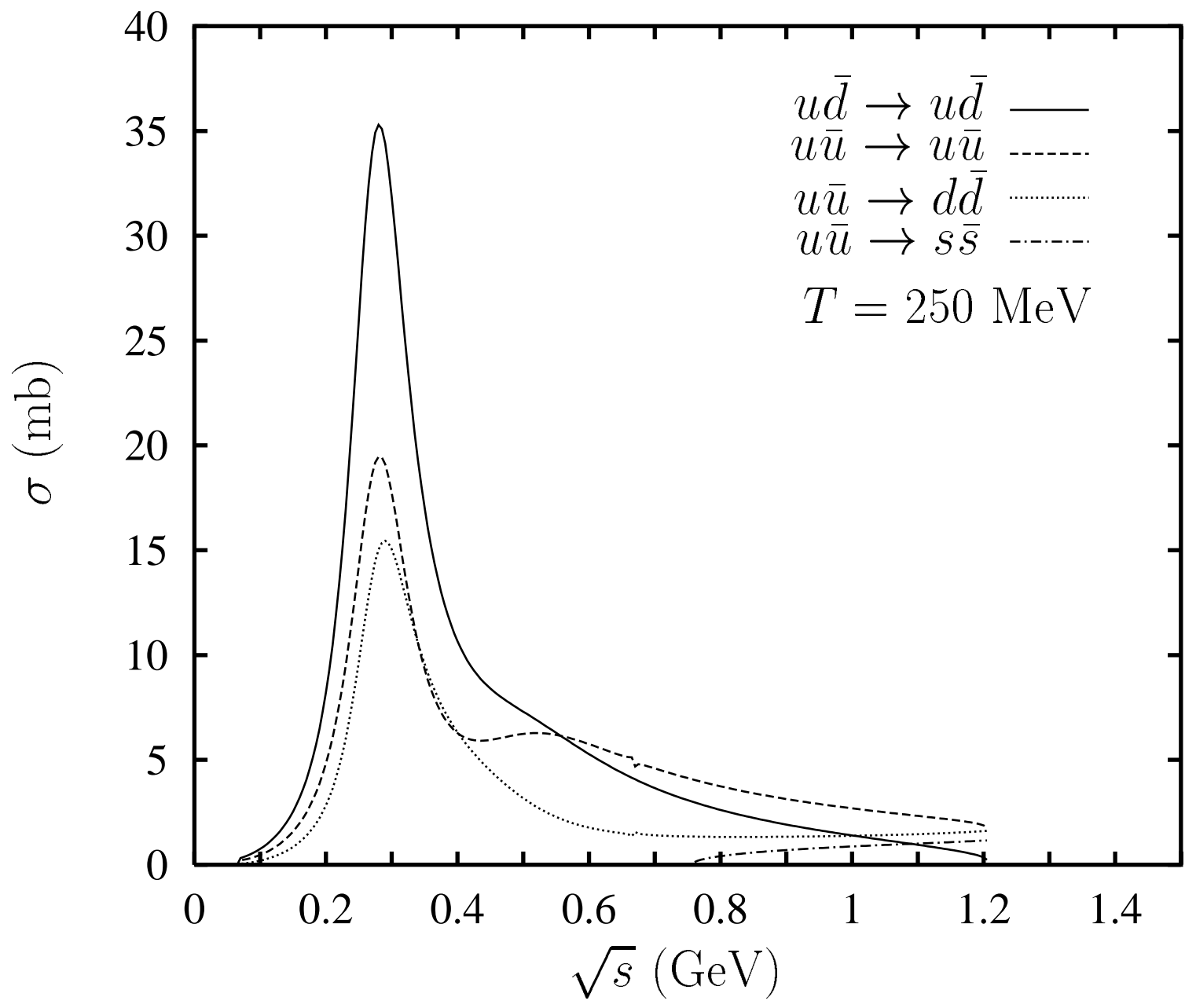
\begin{figure}
\caption[]{Total cross section for $q\bar q$ scattering with only light quarks
	in the initial state, shown as a function of $\sqrt{s}$, at
	$T=250{\rm MeV}$.}
\label{evqqblite}
\end{figure}

In Fig.~\ref{qqb23vergl}, we show a comparison of the scattering cross
sections calculated in $SU_f(3)$ with the corresponding calculation in
$SU_f(2)$, for a specific process $u\bar d\to u\bar d$, at the
temperature $T=215~\mbox{MeV}$. In contrast to the comparison shown in
Fig.~\ref{su23vergl} for quark quark scattering, the difference, on
this scale, is not so great. This is due to the fact that the $q\bar q$
scatterings are resonance dominated:  The temperature chosen lies only
slightly higher than the pionic and kaonic Mott temperatures
$T_{M\pi}=212~\mbox{MeV}$ and $T_{MK}=210~\mbox{MeV}$, so that both
pions and kaons appear as very sharp resonances in the reactions
containing them, see Table~\ref{qqbtab}. We therefore find that large
cross sections occur at threshold for the process shown, in this case
due to the pion resonance. We next show all $q\bar q$ cross sections
that contain no strange quarks or antiquarks in the incoming channel as
a function of $\sqrt{s}$ at the temperature $T=215~\mbox{MeV}$ in
Fig.~\ref{qqblite}. These cross sections all display pronounced
structures, especially in comparison with the quark quark scatterings
of similar flavor (cf.  Fig.~\ref{qqlite}), as they admit resonances in
the $s$--channel.  Cross sections with $u\bar u$ in the initial state
are in addition enhanced by the presence of $\sigma$ mesons. By
contrast, the $u\bar u\to s\bar s$ reaction contains no resonant
structure and remains of the order of 1~mb over its range in
$\sqrt{s}$. At the Mott temperature itself, intermediate states in the
$s$--channel give rise to infinite cross sections at threshold. This
feature, which also appears in other processes like $\pi\pi\to\pi\pi$
\cite{pipi}, $\pi\gamma\to\pi\gamma$ \cite{micki} or $q\bar
q\to\gamma\gamma$ \cite{rost}, is akin to the phenomenon of critical
opalescence and has been discussed in greater detail in
Ref.~\cite{gerry}. For temperatures above the Mott transition, such as
the one shown in this figure, the cross sections are large but finite
at threshold. Increasing the temperature further to $T=250~\mbox{MeV}$,
as has been done for the same set of reactions in Fig.~\ref{evqqblite},
shows that the resonances have become broader and the cross sections in
this regime smaller. Nevertheless, this melting of the pion and sigma
resonances still leads to cross sections that are highly enhanced over
a range of $\sqrt{s}$ between 0.2 and 0.8~GeV in comparison with other
scattering processes, where cross sections of only ${\cal O}
(1~\mbox{mb})$ are found. This remnant of the critical scattering
phenomenon is thus still strongly visible even at high temperatures.
Although it is not possible to observe such scatterings directly in
experiment, it may lead to observable consequences when embedded in a
consistently constructed chiral transport theory.

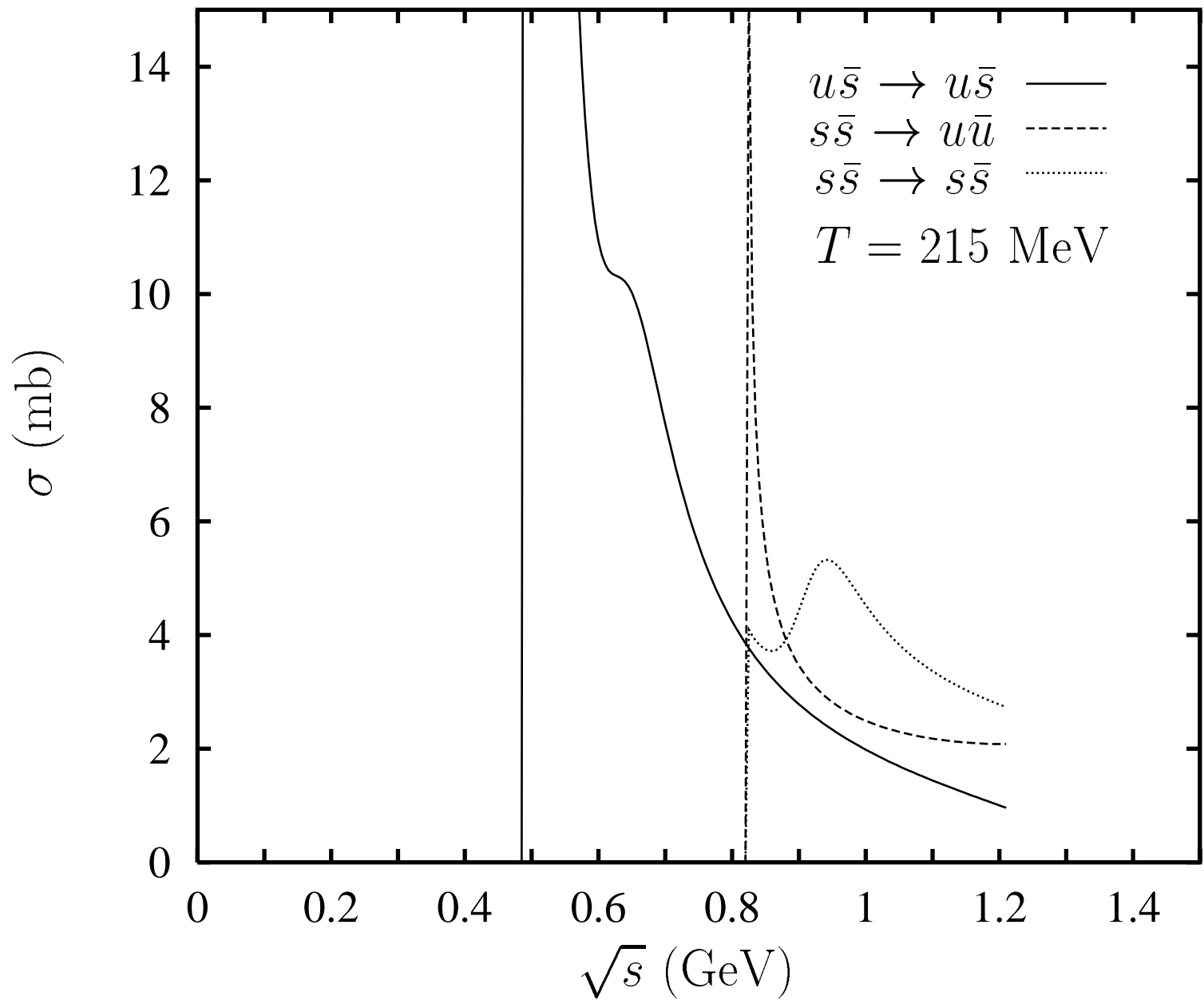
\begin{figure}
\caption[]{Total cross section for $q\bar q$ scattering with at least one
	strange quark in the initial state, shown as a function of
	$\sqrt{s}$, at $T=215{\rm MeV}$.}
\label{qqbstra}
\end{figure}

In Fig.~\ref{qqbstra}, we show the scattering cross sections for quarks
and antiquarks for processes that contain at least one strange particle
in the incoming channel at $T=215~\mbox{MeV}$. The threshold behavior
is strongly dominated by the resonant structure for the case of the
process $u\bar s\to u\bar s$. Here an exchanged kaon gives the dominant
resonance, with a shoulder due to the $\sigma_K$. The $s\bar s\to u\bar
u$ singularity seen is a kinematical singularity, that arises because
the reaction is exothermic. The $s\bar s\to s\bar s$ displays no
pronounced resonance -- there is a small one due to the $\sigma'$ or
$\eta'$. At a somewhat higher temperature, $T=250~\mbox{MeV}$, the
resonant structure that forms a remnant of critical scattering for
$u\bar s\to u\bar s$, has become smaller and broader. Once again, it
still peaks at 15~mb over a range of $\sqrt{s}$ between 0.4 and 1~GeV,
and may therefore also lead to an observable effect when embedded in a
consistent transport theory.

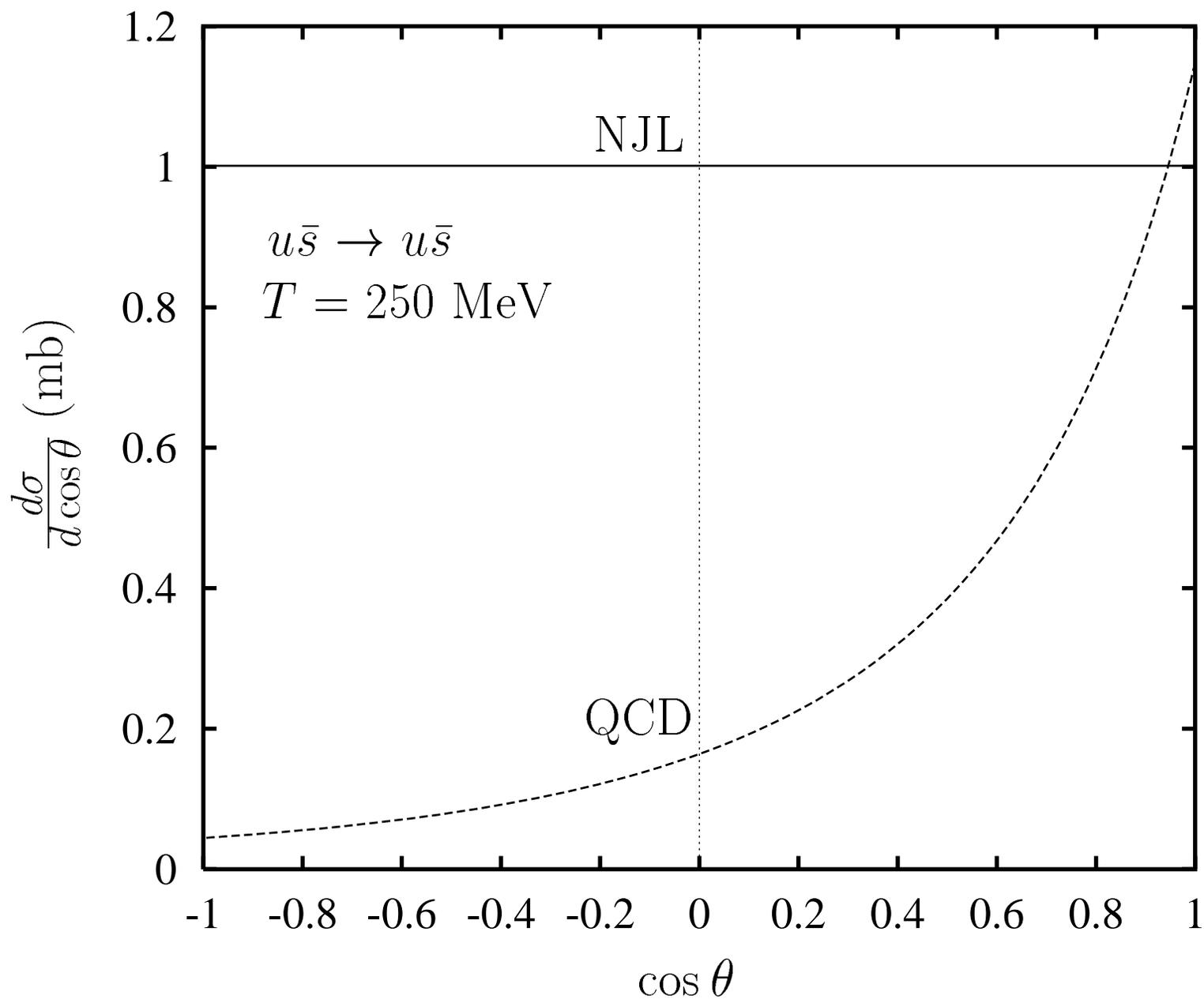
\begin{figure}
\caption[]{Differential cross section for $u\bar s\to u\bar s$ at
	$T=250{\rm MeV}$ and $\sqrt{s}=1~{\rm GeV}$ within the NJL
	model (solid line) and perturbative QCD (dashed line).}
\label{diffqqb}
\end{figure}

Figure~\ref{diffqqb} shows the differential cross section for the
process $u\bar s\to u\bar s$, given as function of $\cos\theta$.
Whereas the one gluon exchange perturbative QCD cross section, see
Appendix~\ref{appqcd}, once again resembles M{\o}ller scattering for
$e^+e^-$ and is strongly peaked in the forward direction, the NJL cross
section is flat. This comes about due to the dominance of the
$s$--channel exchange at this energy, which proceeds via spinless
particles and thus shows no anisotropy.

\section{Transport Properties} \label{secthree}
\subsection{Averaged Transition Rates}
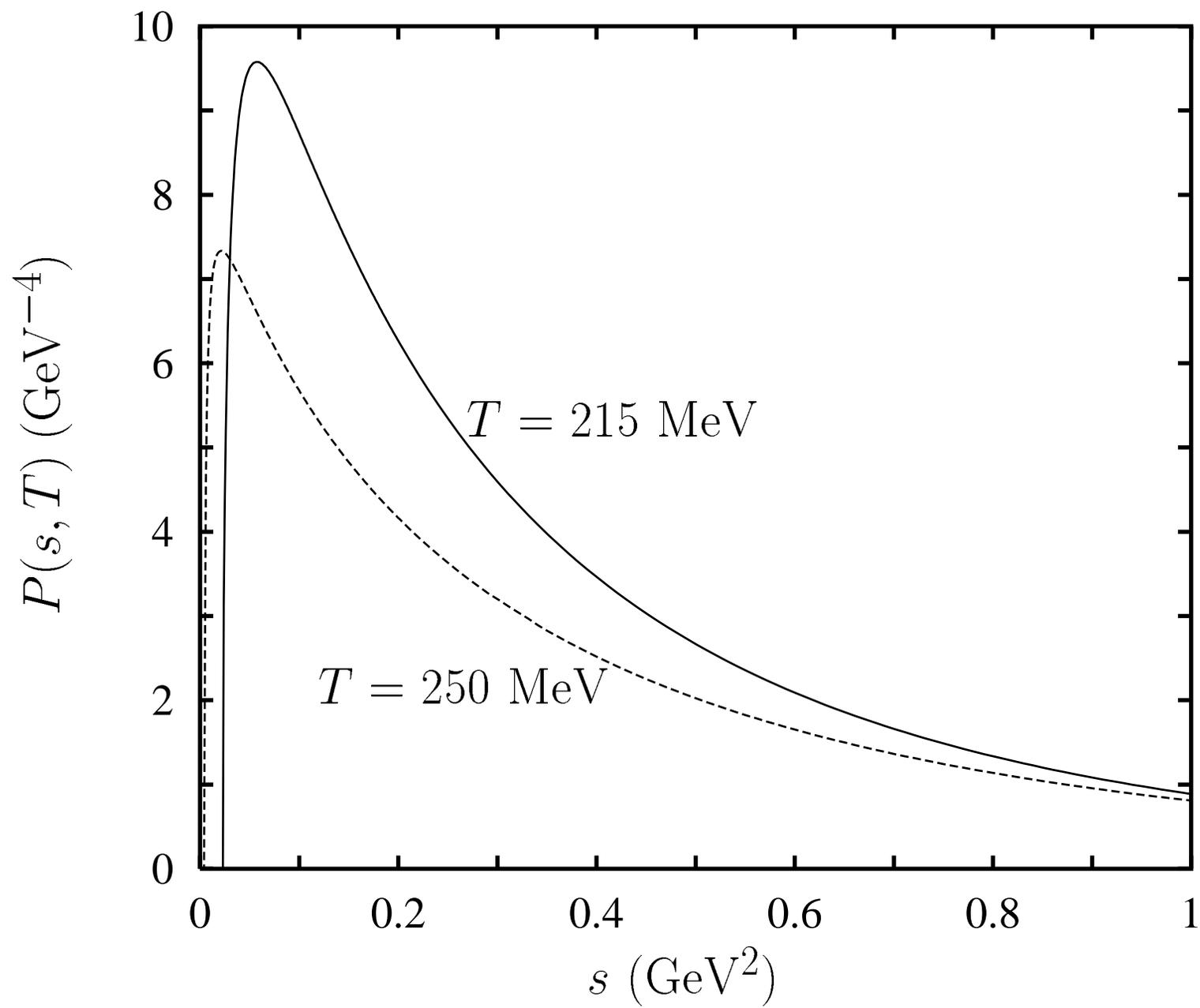
\begin{figure}
\caption[]{Weight function $P(s,T)$ as a function of $s$ for $m_1=m_2=m_u$,
        $T=215~{\rm MeV}$ (solid line) and $T=250~{\rm MeV}$ (dashed line).}
\label{weight}
\end{figure}

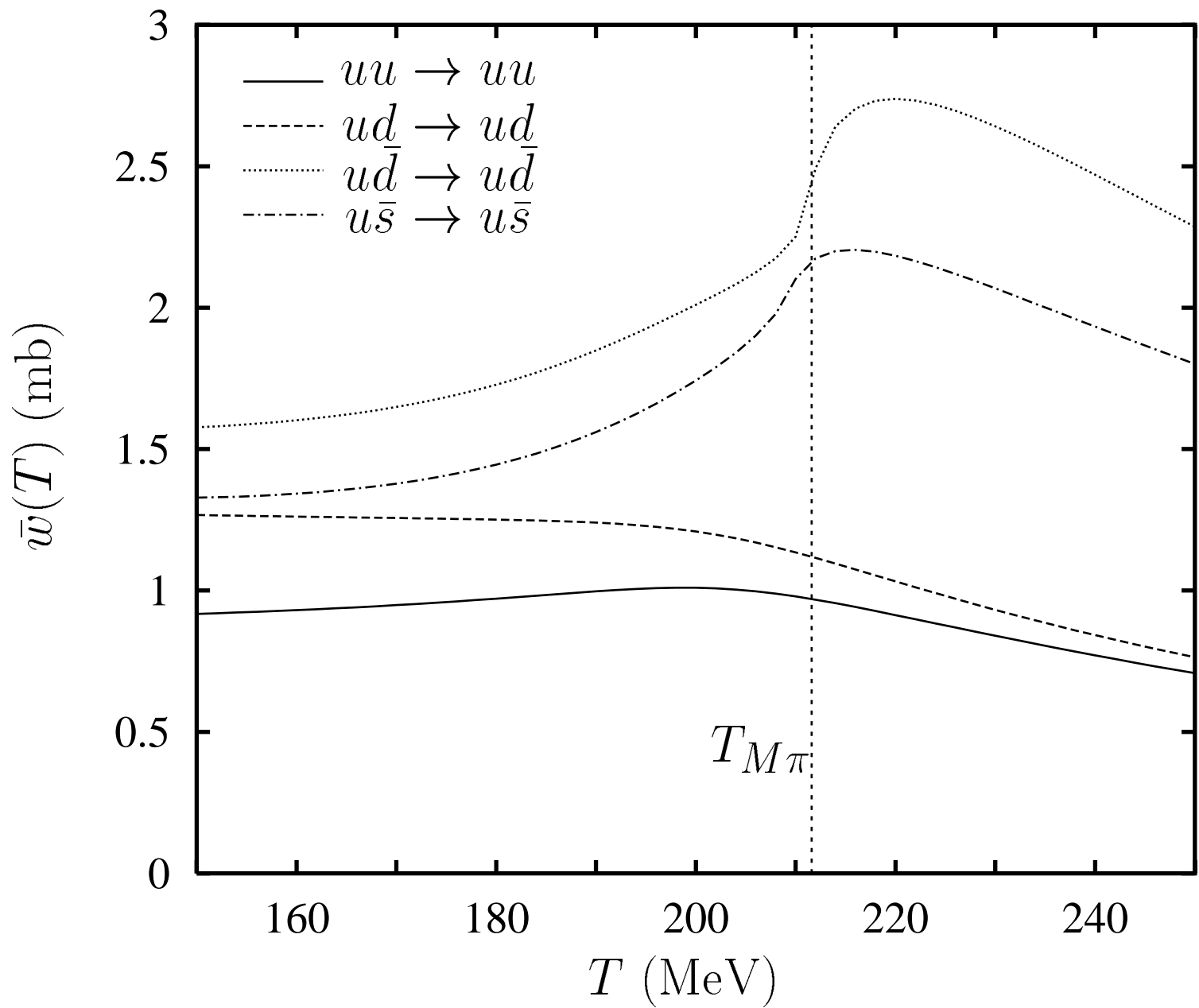
\begin{figure}
\caption[]{Energy averaged transition rates, shown as a function of the
	temperature. The pion Mott temperature is indicated by the
	dashed vertical line.}
\label{avrates}
\end{figure}

To calculate the transport coefficients, we need averaged transition
rates. These are expressed as \cite{su3hadron}
\begin{equation}
\bar w(T) = \frac{1}{\rho_1(T)\rho_2(T)}
\int \frac{d^3p_1}{(2\pi)^3}\frac{d^3p_2}{(2\pi)^3}
\left[2N_cf_F(\beta E_1)\right] \left[2N_cf_F(\beta E_2)\right] w(s,T)
\label{avtrans} \quad,
\end{equation}
in terms of the transition rate $w(s,T)$, which in turn is defined as
the product of cross section and relative velocity:
\begin{equation}
w(s,T)=|\vec v_{\rm rel}| \sigma(s,T)
\end{equation} \begin{equation}
|\vec v_{\rm rel}| = \frac{\sqrt{(p_1p_2)^2-m_1^2m_2^2}}{E_1E_2} \quad.
\end{equation}
The quark density in Eq.~(\ref{avtrans}) is defined as the integral
of the Fermi distribution function over all momenta:
\begin{equation}
\rho_f(T) = \int \frac{d^3p}{(2\pi)^3}
2N_cf_F\left(\beta\sqrt{\vec p^2+m_f^2}\right)
\quad. \label{dichte}
\end{equation}
To evaluate Eq.~(\ref{avtrans}), we make the approximation that the total
cross section is only a function of $s$, even if the incoming pair is
moving with respect to the medium. In this case
the averaged transition rate can be expressed as
\begin{equation}
\bar w(T) = \int_{(m_1+m_2)^2}^\infty ds
\sqrt{(p_1p_2)^2-(m_1m_2)^2} \sigma(s,T) P(s,T) \quad,
\end{equation}
where the weight function $P(s,T)$ is given as
\begin{eqnarray}
P(s,T) &=& \frac{1}{\rho_1(T)\rho_2(T)} \label{pwahr}
\frac{1}{16\pi^4} \int_{m_1}^\infty dE_1 [2N_c f_F(\beta E_1)] \\
& & \quad \times
\int_{m_2}^\infty dE_2
[2N_c f_F(\beta E_2)] \Theta(4|\vec p_1|^2|\vec p_2|^2
-(s-(m_1^2+m_2^2)-2E_1E_2)^2)
\nonumber \quad.
\end{eqnarray}
This weight function is illustrated in Fig.~\ref{weight} for
$m_1=m_2=m_u$, and the two values of the temperature $T=215~{\rm MeV}$
and $T=250~{\rm MeV}$.  One sees that this function weights the lower
energies strongly and has an exponential tail at a given temperature.
Increasing the temperature from $T=215~\mbox{MeV}$ to
$T=250~\mbox{MeV}$ has the consequence that the peak value of the
weight function is shifted and reduced somewhat and the exponential
decay is weaker. Note that this approach differs from that taken in
Ref.~\cite{su2elast} in that $P(s,T)$ in Eq.~(\ref{pwahr}) is {\em
not\/} a normalized probability distribution.  In Fig.~\ref{avrates}, we
show the averaged transition rates for the quark quark scattering
processes $uu\to uu$ and $ud\to ud$ together with the quark antiquark
scattering processes $u\bar d\to u\bar d$ and $u\bar s\to u\bar s$. We
see that the quark antiquark scattering averaged rates lie higher than
the quark quark scattering ones.  There is a sudden rise in the quark
antiquark rates as one moves through the Mott temperatures. This is due
to the inclusion of resonance channels at and above this point.  Above
the Mott temperatures, in the region of interest, both are decreasing
with temperature.

\subsection{Thermal Relaxation Times}
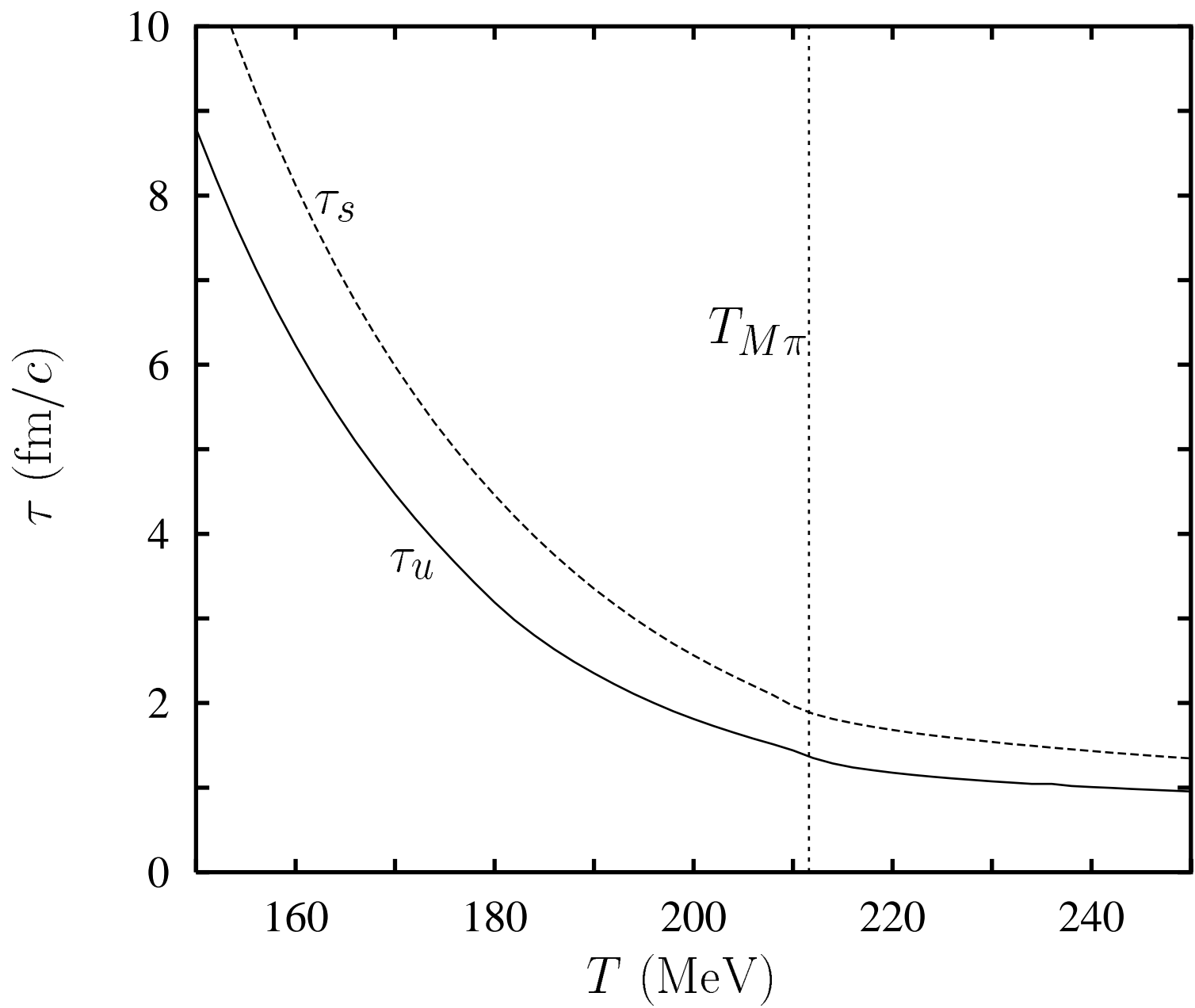
\begin{figure}
\caption[]{Thermal relaxation times for light (solid line) and strange
	 (dashed line) quarks, shown as a function of the temperature.
	The pion Mott temperature is indicated by the dashed vertical
	line.}
\label{relax}
\end{figure}

Since the definition of the averaged transition rate strongly resembles
the collision integral of a Boltzmann equation \cite{deGroot},
we can immediately identify the thermal relaxation time for each
species as
\begin{equation}
\tau_f^{-1} = \sum_{f'} \rho_{f'} \bar w_{ff'} \quad,
\end{equation}
where the summation runs over all quark species and $\bar w_{fg}$ is
the sum of the transition rates of all processes with species $f$ and
$g$ in the initial state. Our numerical calculation for $\tau_u=\tau_d$
and $\tau_s$ is shown as a function of temperature in Fig.~\ref{relax}.
The temperature region below $T_{M\pi}$ is of purely academic interest,
since physically there should be no free quarks: we show it here for
completeness. The relaxation time is large for small temperatures,
which is due to the low quark density. In the physically interesting
region $T\ge T_{M\pi}$, we obtain relaxation times of the order of
1--1.3~fm/$c$ for light quarks and 1.3--2~fm/$c$ for strange quarks
for temperatures greater than the pion Mott temperature.

\subsection{Mean Free Path}
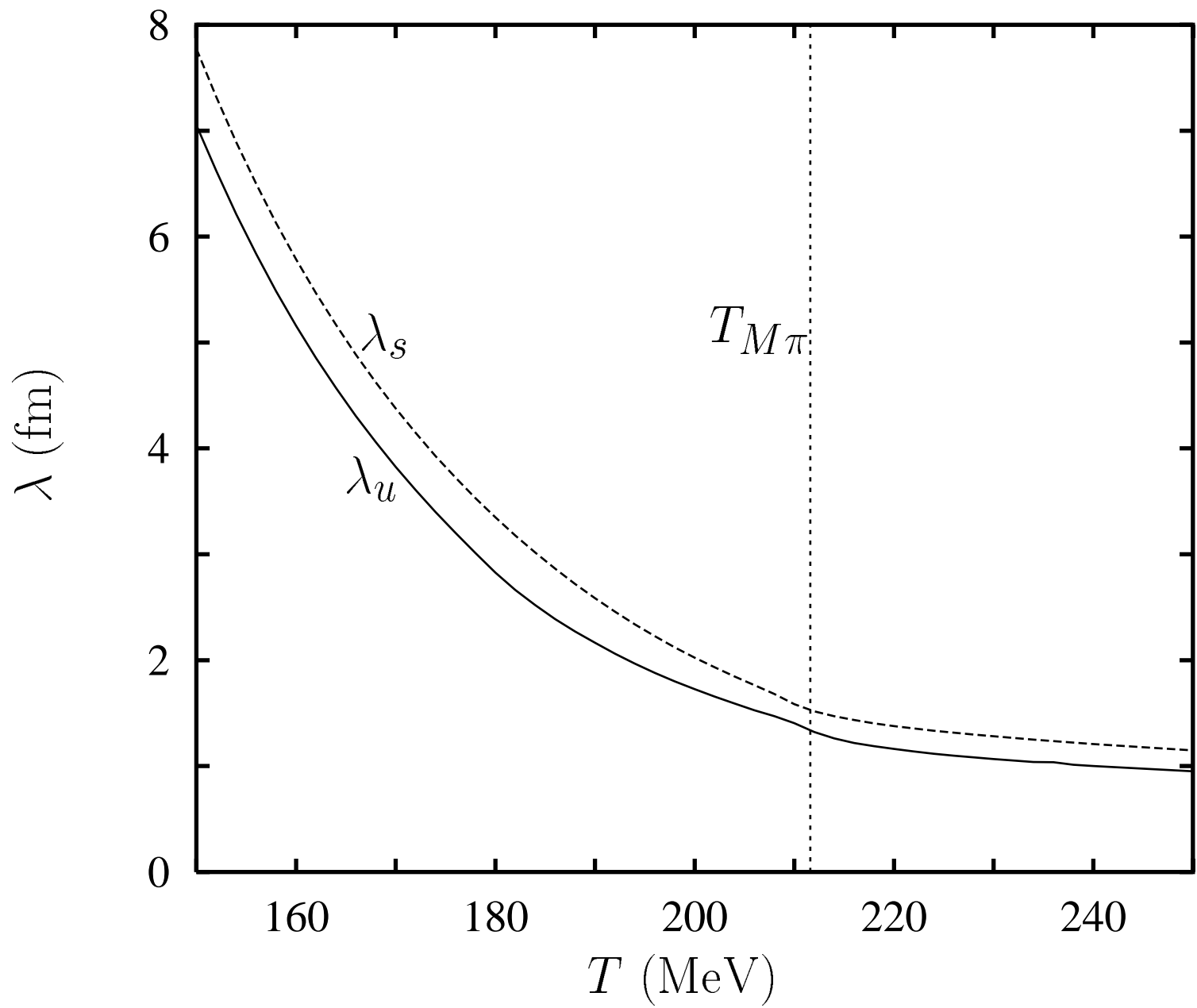
\begin{figure}
\caption[]{Mean free path for light (solid line) and strange
	(dashed line) quarks, shown as a function of the temperature.
	The pion Mott temperature is indicated by the dashed vertical
	line.}
\label{fpath}
\end{figure}

We define the mean free path of a particle as \cite{Reif}
\begin{equation}
\lambda_f = \bar v_f \tau_f \quad,
\end{equation}
where the mean velocity of flavor $f$ is given as
\begin{equation}
\bar v_f = \frac{2 N_c}{\rho_f} \label{vbar}
\int \frac{d^3p}{(2\pi)^3} \frac{p}{E_f} f_F(\beta E_f) \quad.
\end{equation}
Our numerical calculation for $\lambda$ is shown in Fig.~\ref{fpath}.
Once again, we also show the low temperature region, although, as
already pointed out, this is unphysical; one can therefore regard the
physical mean free path as being the value obtained in the high
temperature regime. For high temperatures, $T\gg m_f$, one finds
$\lambda_f\approx\tau_f$, since $\bar v_f$ is approximately equal to
one for all flavors, as can be seen from Eq.~(\ref{vbar}). For lower
temperatures, $\bar v_s<\bar v_u<1$, causing the curves in
Fig.~\ref{fpath} to lie closer together than the corresponding two
curves of Fig.~\ref{relax}.  From Fig.~\ref{fpath}, we obtain a mean
free path of 0.9--1.4~fm for light quarks and 1.1--1.6~fm for strange
quarks.

\subsection{Viscosity}
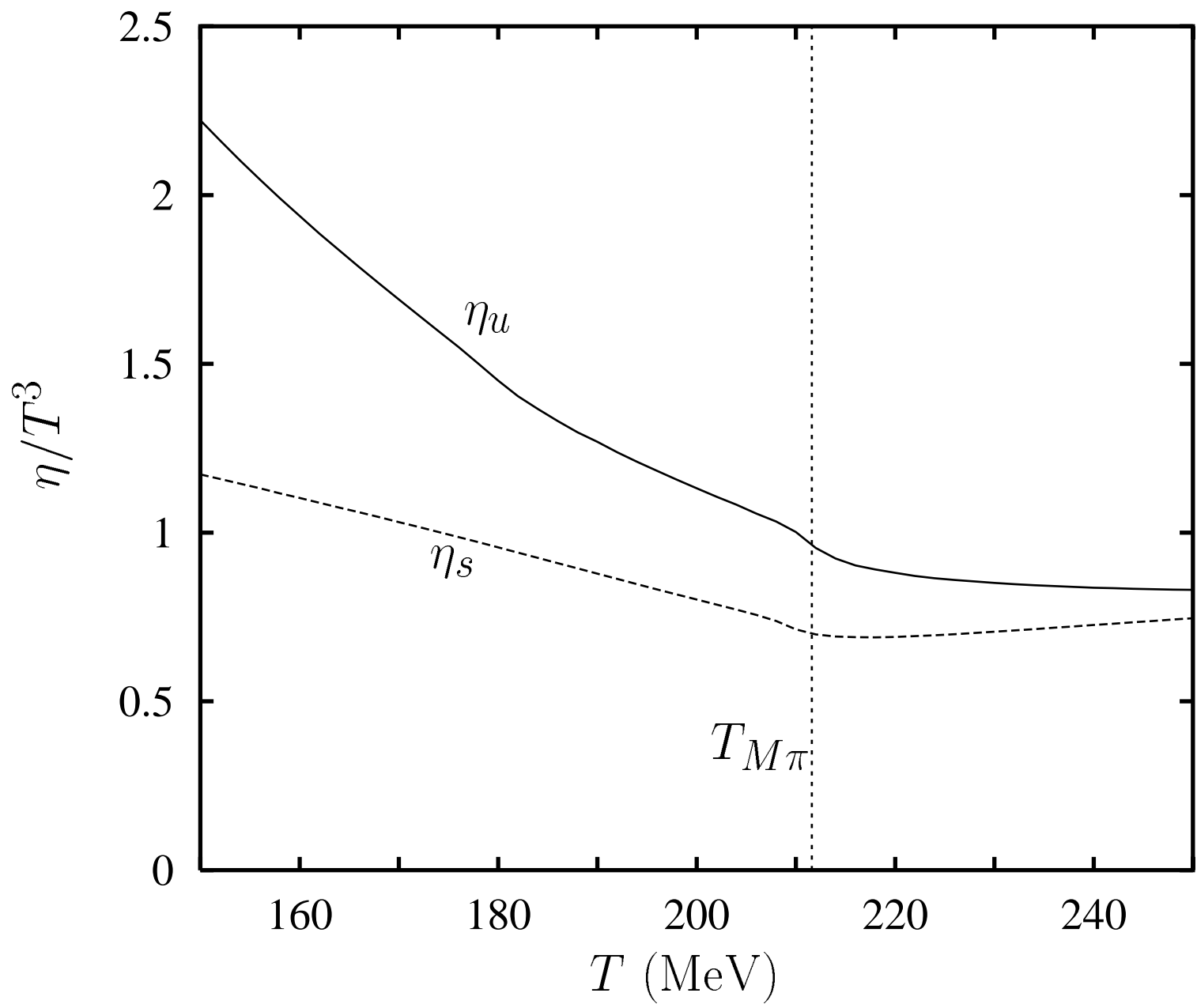
\begin{figure}
\caption[]{Viscosity coefficient per flavor for light (solid line) and
	strange (dashed line) quarks, shown as a function of the
	temperature. The vertical dashed line indicates the pion Mott
	temperature.}
\label{viscofig}
\end{figure}

To a first approximation, the shear viscosity is proportional to the
mean free path \cite{su2elast,Reif,Thoma}:
\begin{equation}
\eta_f = \frac{4}{15} \rho_f \bar p_f \lambda_f \quad, \label{viscos}
\end{equation}
where $\bar p_f$ is the mean momentum for a quark of flavor $f$:
\begin{equation}
\bar p_f = \frac{2 N_c}{\rho_f}
\int \frac{d^3p}{(2\pi)^3} p f_F(\beta E_f) \quad.
\end{equation}
A more detailed analysis \cite{Reif} shows that in averaging the
cross sections, large scattering angles should be preferred, which
is achieved by replacing the total cross section implicitly contained
in Eq.~(\ref{viscos}) with the weighted average
\begin{equation}
\sigma \to \sigma_\eta = \int d\Omega \frac{d\sigma}{d\Omega}
\sin^2 \theta \, [1-f_F(\beta E_3)] [1-f_F(\beta E_4)] \quad.
\end{equation}
Since in a first approximation, $\eta$ should be proportional to $T^3$,
we show $\eta_u/T^3=\eta_d/T^3$ and $\eta_s/T^3$ in Fig.~\ref{viscofig}.
Since the mean free path for all flavors is approximately equal, the
difference in the curves is mainly a density effect. Beyond the Mott
transition, the result for $\eta/T^3$ lies in the range 0.83--1.0
for light quarks and 0.68--0.75 for strange quarks. The total
viscosity, calculated for two flavors, amounts to
\begin{equation}
3.2 < \eta/T^3 < 4.0
\end{equation}
and
\begin{equation}
4.6 < \eta/T^3 < 5.5
\end{equation}
for three flavors. The value for two flavors lies under that given in
Ref.~\cite{su2elast}. This is a direct manifestation of the difference
in the cross sections shown in Fig.~\ref{su23vergl} for two and three
flavors, due to the additional channels that are available in the three
flavor calculation. A classical hydrodynamical estimate \cite{danigu}
places loose bounds on this number,
\begin{equation}
2 \le \eta/T^3 \le 3 \tau T \quad,
\end{equation}
in which $\tau$ is the proper time for the expansion of the system.
For $\tau\approx 5~\mbox{fm}/c$, our values for the total viscosity lie
within these bounds, but are smaller than than those calculated in this
hydrodynamical approach, which gives $\eta/T^3\approx 10$
\cite{danigu}.  Alternatively, perturbative QCD calculations have been
performed \cite{baym,hoka} for this quantity. The variational
perturbative calculation of \cite{baym} places a lower bound on
$\eta$ as being
\begin{equation}
\eta/T^3 \ge - \frac{1.16}{\alpha_s^2\log\alpha_s} \quad.
\end{equation}
For $\alpha_s=0.6$, this is 6.3, which is slightly higher than our
result. Technically this comes about from the fact that the
perturbative QCD calculation favors forward scattering, while the NJL
model allows for a softer angular distribution. The perturbative QCD
result may however be substantially altered in the non--perturbative
region, so we do not regard this as being a serious discrepancy.

\section{Strangeness Enhancement} \label{secfour}
Given the elastic scattering cross sections, we are able to address the
problem of strangeness enhancement in a quark meson plasma within our
model. We choose an approach similar to that taken in
Refs.~\cite{KMR,MueRaf,MSML}.  However, instead of discussing gluon
exchange, we compute the contribution of the exchange of mesonic
resonances.

The strangeness changing processes, which we have considered in this work,
are
\begin{equation}
u\bar u \to s \bar s \hspace{2cm} d\bar d \to s \bar s \label{hinreac}
\end{equation}
and the respective back reactions
\begin{equation}
s\bar s \to u \bar u \hspace{2cm} s\bar s \to d \bar d \quad.
\end{equation}
In the following, we consider the light quarks to be fully equilibrated.
We discuss two possible definitions of the chemical relaxation time.
The first one is defined as the number of strange quarks present in
chemical equilibrium, divided by the number of strange quarks produced
per unit time:
\begin{equation}
\tau_1 = \frac{\rho_s^{\rm eq}}{2 \rho_u \rho_{\bar u}
\bar w_{u\bar u\to s\bar s}} \quad. \label{taupeter}
\end{equation}
The factor 2 in this expression accounts for the number of light quark
flavors, which give equal contributions. This quantity is comparable
with that given in Ref.~\cite{KMR}. In this case one has, as a first
approximation,
\begin{equation}
\rho_s(t) = \rho_s^{\rm eq} \tanh \left(t / \tau_1 \right) \quad.
\end{equation}

The second definition of the strange quark relaxation time is
\begin{equation}
\tau_2 = \frac{1}{2\rho_{\bar u} \bar w_{u\bar u\to s\bar s}}
\quad. \label{taujoerg}
\end{equation}
Again, the factor 2 counts the light quark flavors. This corresponds
to making the rate equation ansatz
\begin{equation}
\frac{d\rho_s}{dt} = \frac{1}{\tau_2}\rho_u \quad,
\end{equation}
neglecting the back reaction. The two definitions of $\tau_1$ and
$\tau_2$ differ by the factor $\rho_s^{\rm eq}(T)/\rho_u(T)$, which is
smaller than one since $m_s(T)>m_u(T)$, but approaches one for
temperatures $T\gg m_s$.

By contrast, in Ref.~\cite{KMR}, the strangeness relaxation time was
found within the framework of perturbative QCD to be dominated by the
process $gg\to s\bar s$. For this process, these authors obtained
\begin{equation}
\tau_{gg} = \frac{1.61}{\alpha_s^2 T}\frac{(m_s/T)\exp(m_s/T)}
{(1 + \frac{99}{56} T/m_s + \dots)} \quad, \label{taukmr}
\end{equation}
where $\alpha_s$ is the strong coupling constant, which was assumed to
be 0.6 for all temperatures.  The relaxation time due to the
processes (\ref{hinreac}) was calculated by the same authors to be
roughly a factor of 10 larger.

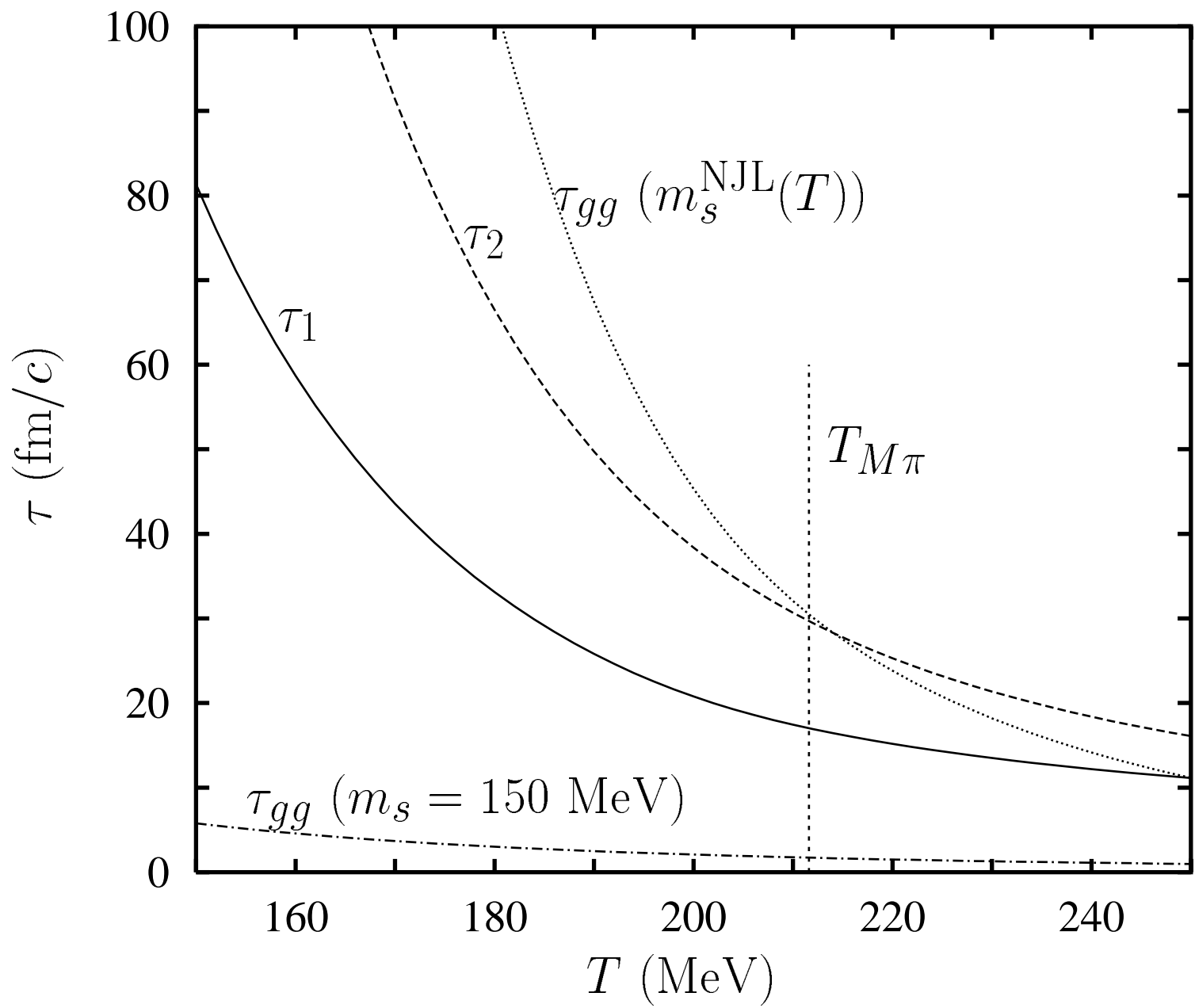
\begin{figure}
\caption[]{Chemical relaxation times for strange quarks, shown as a
	 function of the temperature. The vertical dashed line
	 indicates the pion Mott temperature. The relaxation time
	 calculated according to Eq.~(\ref{taupeter}) is given by the
	 solid line, and according to Eq.~(\ref{taujoerg}) by the
	 dashed line. The dot--dashed line gives the relaxation time
	 calculated according to Eq.~(\ref{taukmr}) with $m_s=150~{\rm
	 MeV}$, while the dotted line gives the relaxation time
	 calculated according to Eq.~(\ref{taukmr}), with strange quark
	 mass from Eq.~(\ref{gapeq}).}
\label{frankie}
\end{figure}

Our numerical calculation of $\tau_1$, $\tau_2$ and $\tau_{gg}$ from
Eqs. (\ref{taupeter}), (\ref{taujoerg}) and (\ref{taukmr}) are shown in
Fig.~\ref{frankie}. Of interest in this figure are the curves in the
temperature range $T>T_{M\pi}$. The solid curve shows $\tau_1$, which
lies in the range 11-17~fm/$c$, whereas $\tau_2$, indicated by the
dashed line, lies in the range 16--30~fm/$c$.  The relaxation time
due to gluon--gluon collisions, as given in Ref.~\cite{KMR}, is given
by the dot--dashed line and is clearly seen to be considerably smaller
than $\tau_1$. However, according to Ref.~\cite{KMR}, this calculation
assumes a temperature independent value for the strange quark mass of
$m_s=150~{\rm MeV}$. The NJL model, on the other hand, predicts a
strange quark mass, which, even at high temperatures, is at least twice
as large as the current quark mass value of 150~MeV \cite{su3hadron}.
In the temperature range shown in the figure,
$150~\mbox{MeV}<T<250~\mbox{MeV}$, the strange quark mass falls from
$510~\mbox{MeV}>m_s>380~\mbox{MeV}$. At the pion Mott temperature
$T_{M\pi}=212~\mbox{MeV}$, we have $m_s=415~\mbox{MeV}$.  An evaluation
of Eq.~(\ref{taukmr}) with masses according to the NJL gap equation
(\ref{gapeq}) is indicated by the dotted curve in Fig.~\ref{frankie}.
Since the chemical relaxation time in Eq.~(\ref{taukmr}) depends
exponentially on $m_s$, an exact knowledge of this quantity is
essential, as can be seen by comparing the dot--dashed and the dotted
curve in Fig.~\ref{frankie}. Taking the strange quark mass from the NJL
gap equation gives a $\tau_{gg}$ of the order of $\tau_1$.
Furthermore, Eq.~(\ref{taukmr}) relies on the assumption, that light
quarks and gluons are massless above the phase transition. Finite light
quark and gluon masses could also shift the chemical relaxation times.

Experimentally, the strangeness content of the observed mesons is
parametrized in the ratio
\begin{equation}
r_{\rm ex} = \frac{4\left<K^0_S\right>}{3\left<\pi^-\right>} \quad,
\end{equation}
where $\left<K^0_S\right>$ and $\left<\pi^-\right>$ are the mean
multiplicities of the observed mesons. At 200~GeV/nucleon the
experimental values are \cite{na35}
\begin{equation}
r_{\rm ex} = \left\{ \begin{array}{ccc}
		( 8.7 \pm 1.3) \% & \qquad & N+N \\
		(15.4 \pm 2.6) \% & \qquad & S+S
\end{array} \right. \quad.
\end{equation}
These values should be compared with the thermal value
\begin{equation}
r_{\rm th} = \frac{\rho_s^{\rm eq}(T)}{\rho_u(T)+\rho_d(T)}=27\%
\end{equation}
at $T=200~\mbox{MeV}$. If we interpret the increase from $r_{\rm
ex}^{N+N}$ to $r_{\rm ex}^{S+S}$ as being due to an ``chemical
equilibration'' within the quark plasma alone, we can determine the
time $t_0$ available for equilibration from the equation
\begin{equation}
\tanh\left(\frac{t_0}{\tau_1}\right) =
\frac{r_{\rm ex}^{S+S}-r_{\rm ex}^{N+N}}{r_{\rm th}} \quad,
\end{equation}
and one obtains $t_0\approx\tau_1/4\approx4~\mbox{fm}/c$. This value is
clearly an upper limit, since strangeness can also be created within
the hadronization process (found to contribute around 1\% in our previous
calculation \cite{su3hadron}) and in the hadronic gas.  This would mean
that one has to assume a plasma lifetime of less than 4~fm$/c$. Note
that calculations using the hydrodynamical Landau model \cite{Udo},
indicate that the total lifetime of the fireball (plasma +
hadronization + hadron gas) is of the order of 5~fm/$c$ for $S+S$
collisions. This is not in contradiction with our estimate.

\section{General Discussion} \label{sechen}
In this section, we discuss general features relating to our
calculation and its philosophy. We have evaluated the quark
self energy in the Hartree approximation, which corresponds to the
leading order term in a $1/N_c$ approximation. The $1/N_c$ expansion is
necessary in order to have a well--defined method of dealing with a strong
coupling theory. Concomitantly, the pion masses are calculated in the
random phase approximation (RPA), which, taken together with the quark
self energy, can be shown to form a consistent expansion in a chiral
sense -- i.e. the Goldstone mode is guaranteed in the chiral limit. The
masses are not shown explicitly here -- the calculation is standard and
has been given in Refs.~\cite{gerry,su3hadron,njlthermo} in detail. In
particular, to this level of approximation, the quark self energy is
always real, even in a thermal medium, while the mesonic self energies
are complex, depending on various parameters.

Given this circumstance, we have calculated quark--quark and
quark--antiquark scattering processes. In a simplistic fashion, by
analogy with the Boltzmann equation, we are able to extract transport
coefficients, and in particular, lifetimes, which have been discussed
in detail in Section~\ref{secthree}.  These lifetimes imply of course that the
$u$, $d$ and $s$ quarks carry widths that are inversely proportional to
these. Thus one sees, that, while this is a consistent chiral and
$1/N_c$ expansion, it is {\em not\/} a self--consistent calculation of
the {\em widths}. In order to do so, one must go {\em beyond\/} the
lowest order $1/N_c$ calculation. Several attempts to do this
\cite{ncexpand} and also to include finite temperature \cite{pengfei}
have difficulties either in including all possible terms, and/or
maintaining chiral symmetry. A first calculation \cite{quackhen} that
attempts to go further and calculate the quark widths
self--consistently ignores
the consistent calculation of the meson sector, and thus violates
chiral symmetry requirements at this level. We find that our widths
$\sim 1/\tau$ are of the
order of 200~MeV and lie somewhat below the values quoted in
Ref.~\cite{quackhen}.

In this discussion, we wish to point out that there are contradicting
expansions that one can examine. It has always been a fixed tenet of
transport theory that truncations should be energy and {\em symmetry\/}
conserving, and we have thus based our calculation on this. Clearly more
work is required to incorporate the higher order terms in a
self--consistent fashion that can also determine quark widths
in medium
simultaneously. As such, we can at best regard our transport theory
results as estimates that may be compared with other similarly made
estimates.

Finally it is perhaps worth commenting that the correct inclusion of
the quasi particles with widths is extremely important. In the
calculation of Ref.~\cite{neise}, the authors have found that a consistent
treatment of the widths leads to a correction of a factor of two for
the collision rates. In addition, a consistent treatment may alter the
transport coefficients substantially, when calculated field
theoretically \cite{hosata,dani}. The imaginary part of the self
energy can also be shown to be connected with memory effects
\cite{henning}.

In concluding this section, it is evident that, within the NJL model,
an expansive study that constructs a transport theory that accounts for
all these problems, still requires substantial development. A first
formal attempt at this has been made in Refs.~\cite{zhawi,wojtek}.

\section{Summary and Conclusions} \label{secfive}
We have examined the finite temperature behavior of elastic scattering
processes of quarks with quarks and quarks with antiquarks within an
effective three flavor model in which chiral symmetry is regarded as
the most important underlying feature. In the quark quark elastic
scattering, we have seen that the three flavor calculation differs
quite substantially from a calculation that accounts for two flavors
only, due to the influence of additional channels. The quark quark
scattering cross sections are enhanced by a factor of 3--4, but are
otherwise featureless. On the other hand the quark antiquark scattering
cross sections display divergencies at the Mott temperature, which we
can regard as an indication of critical scattering. At temperatures
between $T_{M\pi}$ and $T=250~\mbox{MeV}$,
the remnants of the critical scattering, i.e. resonant structures that
occur in the $s$--channel, are clearly visible and enhance the cross
sections substantially in comparison with the background. Such a
feature, it is hoped, may be visible in a final analysis that involves
a consistent transport theory. The differential cross sections that we
have calculated are found to be of the same order of magnitude as
perturbative QCD calculations, in which a gluon mass of
$m_g=600~\mbox{MeV}$ and a strong coupling constant of $\alpha_s=0.6$
have been assumed.

We have also examined transport coefficients in the $SU_f(3)$ model. We
find thermal relaxation times of 1--1.3~fm/$c$ for light quarks and
1.3--2~fm/$c$ for strange quarks. The mean free paths lie in similar
ranges, being 0.9--1.4~fm and 1.1--1.6~fm for light and strange quarks
respectively. The ratios of the viscosity to the third power of the
temperature, $\eta/T^3$ per quark flavor, fall between 0.83--1.0 and
0.68--0.75 for light and strange quarks respectively. The total
viscosity coefficient for two flavors is found to be a factor roughly
two smaller than previous $SU_f(2)$ calculations. This is attributable
to the enhanced cross sections that enter into this $SU_f(3)$
calculation.  Our value still falls within the bracketed range given by
hydrodynamic estimates, while being slightly less than the
perturbative QCD result.

In studying strangeness enhancement due to the possible strangeness
changing reactions that occur via quark antiquark annihilation, we find
that relaxation times are considerably larger than those derived from
perturbative QCD with the naive assumption $m_s=150~\mbox{MeV}$. In the
NJL model, $m_s=m_s(T)$ is a function of temperature and its value lies
significantly higher than the current quark mass value $m_s\approx
150~\mbox{MeV}$ over the entire temperature range of interest. This
feature follows from earlier temperature studies of the dynamically
generated quark masses \cite{su3hadron}.  If one recalculates the
perturbative QCD expression using the temperature dependent NJL quark
masses, one arrives at chemical relaxation times that are of the same
order as our calculation. In attempting to assess the relevance of
elastic scattering processes on strangeness production in heavy ion
collisions, we note that this is a sensitive function of the plasma
lifetime. A plasma lifetime of the order of 4~fm/$c$ would allow one to
account for all of the observed strangeness enhancement in $S+S$
collisions via the mechanism of elastic scattering. Since the plasma
lifetime probably lies below this value, other processes such as
hadronization and final state interactions must also play a role.

\section*{Acknowledgments}
We wish to thank P.~Zhuang and D.~Blaschke for illuminating
discussions.  This work has been supported in part by the Deutsche
Forschungsgemeinschaft under contract no. Hu~233/4--3, by the Federal
Ministry for Education and Research, under contract no. 06~HD~742.

\begin{appendix}
\section{Squared Transition Amplitudes} \label{ampliapp}
\subsection{Quark Quark Scattering}
The matrix elements in Eqs.~(\ref{tchan}) and (\ref{uchan}) have the
form
\begin{eqnarray}
-i{\cal M}_t &=& \delta_{c_1,c_3} \delta_{c_2,c_4}
                 \bar u(p_3)u(p_1)
                 \left[i{\cal D}_t^S(p_1-p_3)\right]
                 \bar u(p_4) u(p_2)
\\ \nonumber
             &+& \delta_{c_1,c_3} \delta_{c_2,c_4}
                 \bar u(p_3)(i\gamma_5)u(p_1)
                 \left[i{\cal D}_t^P(p_1-p_3)\right]
                 \bar u(p_4)(i\gamma_5)u(p_2)
\end{eqnarray}
and
\begin{eqnarray}
-i{\cal M}_u &=& \delta_{c_1,c_4} \delta_{c_2,c_3}
                 \bar u(p_4)u(p_1)
                 \left[i{\cal D}_u^S(p_1-p_4)\right]
                 \bar u(p_3)u(p_2)
\\ \nonumber
             &+& \delta_{c_1,c_4} \delta_{c_2,c_3}
                 \bar u(p_4)(i\gamma_5)u(p_1)
                 \left[i{\cal D}_u^P(p_1-p_3)\right]
                 \bar u(p_3)(i\gamma_5)u(p_2) \quad.
\end{eqnarray}
Here we have dropped the $T$ factors for simplicity, since they
can easily be included by a rescaling of the propagators. After a
short calculation, one obtains
\begin{eqnarray}
\frac{1}{4N_c^2}\sum_{s,c} \left| {\cal M}_t \right|^2 &=&
\frac{1}{4}\Big\{ \left|{\cal D}_t^S \right|^2
{\rm tr} [(p_3\!\!\!\!\!/\,+m_3)(p_1\!\!\!\!\!/\,+m_1)]
{\rm tr}[(p_4\!\!\!\!\!/\,+m_4)(p_2\!\!\!\!\!/\,+m_2)]
\\ \nonumber &+&
\left|{\cal D}_t^P \right|^2
{\rm tr}[(p_3\!\!\!\!\!/\,+m_3) \gamma_5 (p_1\!\!\!\!\!/\,+m_1) \gamma_5]
{\rm tr}[(p_4\!\!\!\!\!/\,+m_4) \gamma_5 (p_2\!\!\!\!\!/\,+m_2) \gamma_5] \Big\}
\quad.
\end{eqnarray}
The interference term between the scalar and pseudoscalar part vanishes.
The final result can be expressed using the Mandelstam variable $t$:
\begin{equation}
\frac{1}{4N_c^2}\sum_{s,c} \left| {\cal M}_t \right|^2 =
\left|{\cal D}_t^S \right|^2 t_{13}^+ t_{24}^+ +
\left|{\cal D}_t^P \right|^2 t_{13}^- t_{24}^- \quad, \label{A4}
\end{equation}
where we have abbreviated $t_{ij}^\pm = t - (m_i \pm m_j)^2$.
Analogously, one obtains
\begin{equation}
\frac{1}{4N_c^2}\sum_{s,c} \left| {\cal M}_u \right|^2 =
\left|{\cal D}_u^S \right|^2 u_{14}^+ u_{23}^+ +
\left|{\cal D}_u^P \right|^2 u_{14}^- u_{23}^- \quad, \label{A5}
\end{equation}
with $u_{ij}^\pm = u - (m_i \pm m_j)^2$.
For the interference term, the spin summation results in
\begin{eqnarray}
\frac{1}{4N_c^2}\sum_{s,c}{\cal M}_t {\cal M}_u^* &=&
\frac{1}{4N_c} \Big\{ {\cal D}_t^S {\cal D}_u^{S*}
{\rm tr}[(p_3\!\!\!\!\!/\,+m_3)(p_1\!\!\!\!\!/\,+m_1)
         (p_4\!\!\!\!\!/\,+m_4)(p_2\!\!\!\!\!/\,+m_2)]
\label{A6} \\ \nonumber &-&
{\cal D}_t^S {\cal D}_u^{P*}
{\rm tr}[(p_3\!\!\!\!\!/\,+m_3)(p_1\!\!\!\!\!/\,+m_1) \gamma_5
         (p_4\!\!\!\!\!/\,+m_4)(p_2\!\!\!\!\!/\,+m_2) \gamma_5]
\\ \nonumber &-&
{\cal D}_t^P {\cal D}_u^{S*}
{\rm tr}[(p_3\!\!\!\!\!/\,+m_3) \gamma_5 (p_1\!\!\!\!\!/\,+m_1)
         (p_4\!\!\!\!\!/\,+m_4) \gamma_5 (p_2\!\!\!\!\!/\,+m_2)]
\\ \nonumber &+&
{\cal D}_t^P {\cal D}_u^{P*}
{\rm tr}[(p_3\!\!\!\!\!/\,+m_3) \gamma_5 (p_1\!\!\!\!\!/\,+m_1) \gamma_5
         (p_4\!\!\!\!\!/\,+m_4) \gamma_5 (p_2\!\!\!\!\!/\,+m_2) \gamma_5] \Big\}
\\ \nonumber &=&
\frac{1}{4N_c} \Big[
{\cal D}_t^S {\cal D}_u^{S*}
\left( t_{13}^+ t_{24}^+ - s_{12}^+ s_{34}^+ + u_{14}^+ u_{23}^+ \right)
\\ \nonumber &-&
{\cal D}_t^S {\cal D}_u^{P*}
\left( t_{13}^+ t_{24}^+ - s_{12}^- s_{34}^- + u_{14}^- u_{23}^- \right)
\\ \nonumber &-&
{\cal D}_t^P {\cal D}_u^{S*}
\left( t_{13}^- t_{24}^- - s_{12}^- s_{34}^- + u_{14}^+ u_{23}^+ \right)
\\ \nonumber &+&
{\cal D}_t^P {\cal D}_u^{P*}
\left( t_{13}^- t_{24}^- - s_{12}^+ s_{34}^+ + u_{14}^- u_{23}^- \right)
\Big] \quad.
\end{eqnarray}
Here, we have abbreviated $s_{ij}^\pm = s - (m_i\pm m_j)^2$.  If the
masses of all incoming and outgoing particles are equal,
Eqs.~(\ref{A4})--(\ref{A6}) can be largely simplified to yield
\begin{eqnarray}
\frac{1}{4N_c^2}\sum_{s,c} \left| {\cal M}_t \right|^2 &=&
\left|{\cal D}_t^S \right|^2 (t-4m^2)^2 + \left|{\cal D}_t^P \right|^2 t^2
\quad, \\ \frac{1}{4N_c^2}\sum_{s,c} \left| {\cal M}_u \right|^2 &=&
\left|{\cal D}_u^S \right|^2 (u-4m^2)^2 + \left|{\cal D}_u^P \right|^2 u^2
\quad,
\end{eqnarray} and \begin{eqnarray}
\frac{1}{4N_c^2}\sum_{s,c}{\cal M}_t {\cal M}_u^* &=&
-\frac{1}{2N_c} \Big\{
{\cal D}_t^S {\cal D}_u^{S*} [tu + 4m^2 (t+u) - 16 m^4]
\\ \nonumber &-&
{\cal D}_t^S {\cal D}_u^{P*} u ( t - 4m^2)
- {\cal D}_t^P {\cal D}_u^{S*} t (u - 4m^2)
+ {\cal D}_t^P {\cal D}_u^{P*} tu \Big\} \quad,
\end{eqnarray}
which is the result of Ref.~\cite{su2elast}.

\subsection{Quark Antiquark Scattering}
The matrix elements for $q\bar q$ scattering are
\begin{eqnarray}
-i{\cal M}_s &=& \delta_{c_1,c_2} \delta_{c_3,c_4}
                 \bar v(p_2) u(p_1)
                 \left[i{\cal D}_s^S(p_1+p_2)\right]
                 \bar u(p_3) v(p_4)
\\ \nonumber
             &+& \delta_{c_1,c_2} \delta_{c_3,c_4}
                 \bar v(p_2) (i\gamma_5) u(p_1)
                 \left[i{\cal D}_s^P(p_1+p_2)\right]
                 \bar u(p_3) (i\gamma_5) v(p_4)
\end{eqnarray}
and
\begin{eqnarray}
-i{\cal M}_t &=& \delta_{c_1,c_3} \delta_{c_2,c_4}
                 \bar u(p_4) u(p_1)
                 \left[i{\cal D}_t^S(p_1-p_3)\right]
                 \bar v(p_3) v(p_2)
\\ \nonumber
             &+& \delta_{c_1,c_3} \delta_{c_2,c_4}
                 \bar u(p_4) (i\gamma_5) u(p_1)
                 \left[i{\cal D}_t^P(p_1-p_3)\right]
                 \bar v(p_3) (i\gamma_5) v(p_2) \quad.
\end{eqnarray}
The square of these matrix elements can be immediately written down
by using the fact that the crossing transformation
\begin{equation}
\begin{array}{lll}
p_1 \to p_1  & \qquad & p_2 \to -p_4 \\
p_3 \to -p_2 & \qquad & p_4 \to p_3 \\
\end{array}
\end{equation}
transforms the $t$ ($u$) channel of $qq$ scattering to the $s$ ($t$)
channel for $q\bar q$ scattering. Thus one obtains
\begin{eqnarray}
\frac{1}{4N_c^2}\sum_{s,c} \left| {\cal M}_s \right|^2 &=&
\left|{\cal D}_s^S \right|^2 s_{12}^+ s_{34}^+ +
\left|{\cal D}_s^P \right|^2 s_{12}^- s_{34}^- \quad, \\
\frac{1}{4N_c^2}\sum_{s,c} \left| {\cal M}_t \right|^2 &=&
\left|{\cal D}_t^S \right|^2 t_{13}^+ t_{24}^+ +
\left|{\cal D}_t^P \right|^2 t_{13}^- t_{24}^- \quad,
\end{eqnarray} and \begin{eqnarray}
\frac{1}{4N_c^2}\sum_{s,c}{\cal M}_s {\cal M}_t^* &=&
\frac{1}{4N_c} \Big[
{\cal D}_s^S {\cal D}_t^{S*}
\left( s_{12}^+ s_{34}^+ - u_{14}^+ u_{24}^+ + t_{13}^+ t_{24}^+ \right)
\\ \nonumber &-&
{\cal D}_s^S {\cal D}_t^{P*}
\left( s_{12}^+ s_{34}^+ - u_{14}^- u_{24}^- + t_{13}^- t_{24}^- \right)
\\ \nonumber &-&
{\cal D}_s^P {\cal D}_t^{S*}
\left( s_{12}^- s_{34}^- - u_{14}^- u_{24}^- + t_{13}^+ t_{24}^+ \right)
\\ \nonumber &+&
{\cal D}_s^P {\cal D}_t^{P*}
\left( s_{12}^- s_{34}^- - u_{14}^+ u_{24}^+ + t_{13}^- t_{24}^- \right)
\Big] \quad.
\end{eqnarray}
Again, these expressions can be greatly simplified in the case of equal
quark masses.

\section{Elastic Scattering in Perturbative QCD} \label{appqcd}
In perturbative QCD, the elastic scattering of quarks and antiquarks
proceeds in lowest order via one gluon exchange. A previous calculation
for the case of massless quarks and gluons has been given in
Ref.~\cite{cusie}. Here we extend this result to the case of three
flavors and introduce finite masses for quarks and gluons.

\subsection{Quark Quark Scattering}
The elastic scattering of two quarks of flavor $f$ and $f'$ proceeds
via a $t$--channel exchange, if $f\ne f'$, and via $t$-- and $u$--channel
exchanges if $f=f'$. The differential cross section is
\begin{equation}
\frac{d\sigma}{dt} = \frac{1}{16\pi[s-(m_f-m_{f'})^2][s-(m_f+m_{f'})^2]}
\frac{1}{4N_c^2}\sum_{s,c}\left|{\cal M}_t-\delta_{f,f'}{\cal M}_u\right|^2
\quad.
\end{equation}
For the $t$--channel exchange, one has, using the Feynman gauge,
\begin{equation}
-i{\cal M}_t= \bar u(p_3)\left(-ig_s\gamma^\mu\frac{1}{2}\lambda^a_{ij}\right)
               u(p_1) \frac{-i\delta_{ab}g_{\mu\nu}}{t-m_g^2} \bar u(p_4)
              \left(-ig_s\gamma^\nu\frac{1}{2}\lambda^b_{kl}\right) u(p_2)
\quad,
\end{equation}
where we have introduced a finite gluon mass $m_g$ in order to avoid
Coulomb singularities. For the $u$-channel, one obtains
\begin{equation}
-i{\cal M}_u= \bar u(p_3)\left(-ig_s\gamma^\mu\frac{1}{2}\lambda^a_{il}\right)
               u(p_2) \frac{-i\delta_{ab}g_{\mu\nu}}{u-m_g^2} \bar u(p_4)
              \left(-ig_s\gamma^\nu\frac{1}{2}\lambda^b_{kj}\right) u(p_1)
\quad.
\end{equation}
Using standard techniques, the expression
\begin{eqnarray}
\frac{1}{4N_c^2}\sum_{s,c}\left|{\cal M}_t-\delta_{f,f'}{\cal M}_u\right|^2
&=&\frac{64\pi^2\alpha_s^2}{9}\Bigg[
\frac{2(s-m_f^2-m_{f'}^2)^2+t^2+2st}{(t-m_g^2)^2}
\label{qmast} \\ &+& \nonumber
\delta_{f,f'}\frac{2(s-2m_f^2)^2+u^2+2su}{(u-m_g^2)^2}
\\ &-& \nonumber
\frac{2}{3}\delta_{f,f'}\frac{(s-4m_f^2)^2-4m_f^4}{(t-m_g^2)(u-m_g^2)}
\Bigg]
\end{eqnarray}
can be derived, where $\alpha_s=g_s^2/(4\pi)$ is the QCD fine structure
constant.

\subsection{Quark Antiquark Scattering}
In the case of different incoming flavors, elastic scattering proceeds
via a $t$--channel exchange. In this case, we have
\begin{equation}
\frac{d\sigma}{dt} = \frac{1}{16\pi[s-(m_f-m_{f'})^2][s-(m_f+m_{f'})^2]}
\frac{1}{4N_c^2}\sum_{s,c}\left|{\cal M}_t\right|^2 \quad,
\end{equation}
and the square of the scattering amplitude can be obtained from the
$t$--channel amplitude contained in Eq.~(\ref{qmast}) by swapping $s$
and $u$:
\begin{equation} \label{smast}
\frac{1}{4N_c^2}\sum_{s,c}\left|{\cal M}_t\right|^2
=\frac{64\pi^2\alpha_s^2}{9}\left[
\frac{2(u-m_f^2-m_{f'}^2)^2+t^2+2ut}{(t-m_g^2)^2} \right] \quad.
\end{equation}
In the case of equal incoming flavors, we have to distinguish two
possibilities: (i) the incoming and outgoing pair have equal flavor,
and (ii) the incoming and outgoing pair have different flavor.
Whereas the second process proceeds via an $s$--channel only,
the first one also has a $t$--channel available.
We now switch our notation to denote the flavor of the
incoming pair by $f$ and the flavor of the outgoing pair by $f'$.
The differential cross section then takes the form
\begin{equation}
\frac{d\sigma}{dt} = \frac{1}{16\pi s(s-4m_f)^2}
\frac{1}{4N_c^2}\sum_{s,c}\left|{\cal M}_s-\delta_{f,f'}{\cal M}_t\right|^2
\quad.
\end{equation}
The squared transition amplitude can be obtained from Eq.~(\ref{qmast})
by making the substitution $s\to u\to t\to s$, i.e. one has
\begin{eqnarray} \label{tmast}
\frac{1}{4N_c^2}\sum_{s,c}\left|{\cal M}_s-\delta_{f,f'}{\cal M}_t\right|^2
&=&\frac{64\pi^2\alpha_s^2}{9}\Bigg[
\frac{2(u-m_f^2-m_{f'}^2)^2+s^2+2su}{(s-m_g^2)^2}
\\ \nonumber &+&
\delta_{f,f'}\frac{2(u-2m_f^2)^2+t^2+2tu}{(t-m_g^2)^2}
\\ \nonumber &-&
\frac{2}{3}\delta_{f,f'}\frac{(u-4m_f^2)^2-4m_f^4}{(s-m_g^2)(t-m_g^2)}
\Bigg] \quad.
\end{eqnarray}
In the massless limit, Eqs.~(\ref{qmast}), (\ref{smast}) and (\ref{tmast})
reduce to the result of Ref.~\cite{cusie}.

\end{appendix}

\begin{table}
\caption[]{Independent processes for $qq$ scattering.} \vspace{2mm}
\begin{tabular}{|c||c|c|}
Process & Exchanged Mesons in $t$ Channel & Exchanged Mesons in $u$ Channel \\
\hline
$uu\to uu$ & $\pi$, $\eta$, $\eta'$, $\sigma_\pi$, $\sigma$, $\sigma'$ &
             $\pi$, $\eta$, $\eta'$, $\sigma_\pi$, $\sigma$, $\sigma'$ \\
$ss\to ss$ & $\eta$, $\eta'$, $\sigma$, $\sigma'$ &
             $\eta$, $\eta'$, $\sigma$, $\sigma'$ \\
$ud\to ud$ & $\pi$, $\eta$, $\eta'$, $\sigma_\pi$, $\sigma$, $\sigma'$ &
             $\pi$, $\sigma_\pi$ \\
$us\to us$ & $\eta$, $\eta'$, $\sigma$, $\sigma'$ &
             $K$, $\sigma_K$ \\
\end{tabular}
\label{qqtab}
\end{table}

\begin{table}
\caption[]{Independent processes for $q\bar q$ scattering.} \vspace{2mm}
\begin{tabular}{|c||c|c|}
Process & Exchanged Mesons in $s$ Channel & Exchanged Mesons in $t$ Channel \\
\hline
$u\bar d \to u \bar d$ & $\pi$, $\sigma_\pi$
                       & $\pi$, $\eta$, $\eta'$, $\sigma_\pi$, $\sigma$,
                         $\sigma'$ \\
$u\bar s \to u \bar s$ & $K$, $\sigma_K$
                       & $\eta$, $\eta'$, $\sigma$, $\sigma'$ \\
$u\bar u \to u \bar u$ & $\pi$, $\eta$, $\eta'$, $\sigma_\pi$, $\sigma$,
                         $\sigma'$
                       & $\pi$, $\eta$, $\eta'$, $\sigma_\pi$, $\sigma$,
                         $\sigma'$ \\
$u\bar u \to d \bar d$ & $\pi$, $\eta$, $\eta'$, $\sigma_\pi$, $\sigma$,
                         $\sigma'$
                       & $\pi$, $\sigma_\pi$ \\
$u\bar u \to s \bar s$ & $\eta$, $\eta'$, $\sigma$, $\sigma'$
                       & $K$, $\sigma_K$ \\
$s\bar s \to u \bar u$ & $\eta$, $\eta'$, $\sigma$, $\sigma'$
                       & $K$, $\sigma_K$ \\
$s\bar s \to s \bar s$ & $\eta$, $\eta'$, $\sigma$, $\sigma'$
                       & $\eta$, $\eta'$, $\sigma$, $\sigma'$ \\
\end{tabular}
\label{qqbtab}
\end{table}

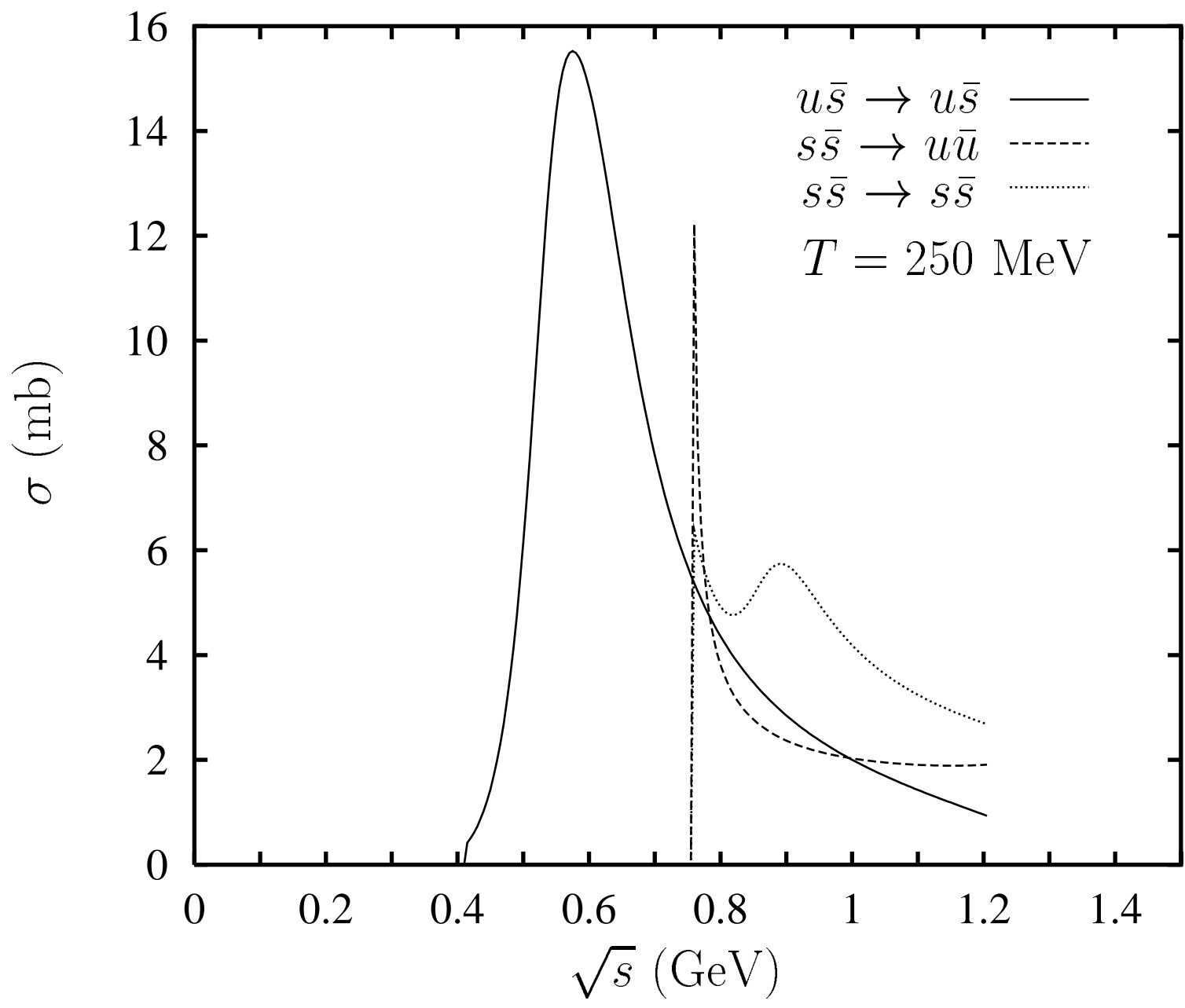
\begin{figure}
\caption[]{Total cross section for $q\bar q$ scattering with at least one
	strange quark in the initial state, shown as a function of
	$\sqrt{s}$, at $T=250{\rm MeV}$.}
\label{evqqbstra}
\end{figure}
 
\begin{references}
\bibitem{heinz} U.~Heinz, Phys. Rev. Lett. 51 (1983) 351; Ann. Phys.
        (NY) 161 (1985) 48.
\bibitem{elze} H.~Th.~Elze, M.~Gyulassy and D.~Vasak, Nucl. Phys. B 276
        (1986) 706; Phys. Lett. B 177 (1986) 402.
\bibitem{vogl} U.~Vogl and W.~Weise, Prog. Part. Nucl. Phys. 27
        (1991) 195.
\bibitem{sandi} S.~P.~Klevansky, Rev. Mod. Phys. 64 (1992) 649.
\bibitem{hatkun} T.~Hatsuda and T.~Kunihiro, Phys. Rep. 247 (1994) 221.
\bibitem{gerry} J.~H\"ufner, S.~P.~Klevansky and P.~Rehberg, University of
        Heidelberg Report No.~HD--TVP--96--03 (1996), to appear in
        Physics Reports.
\bibitem{ncexpand} E.~Quack and S.~P.~Klevansky, Phys. Rev C 49
        (1994) 3283; J.~M\"uller and S.~P.~Klevansky, Phys. Rev. C 50
        (1994) 410; V.~Dmitra\v sinovi\' c, H.~J.~Schulze, R.~Tegen and
        R.~H.~Lemmer, Ann. Phys. (NY) 238 (1995) 332.
\bibitem{su3hadron} P.~Rehberg, S.~P.~Klevansky and J.~H\"ufner,
        Phys. Rev. C 53 (1996) 410.
\bibitem{su2elast} P.~Zhuang, J.~H\"ufner, S.~P.~Klevansky and L.~Neise,
        Phys. Rev. D 51 (1995) 3728.
\bibitem{baym} G.~Baym, H.~Monien, C.~J.~Pethick and D.~G.~Ravenhall,
        Phys. Rev. Lett. 64 (1990) 1867.
\bibitem{hoka} A.~Hosoya and K.~Kajantie, Nucl. Phys. B 250 (1985) 666.
\bibitem{danigu} P.~Danielewicz and M.~Gyulassy, Phys. Rev. D 31
        (1985) 53.
\bibitem{KMR} P.~Koch, B.~M\"uller and J.~Rafelski, Phys. Rep. 142
        (1986) 167.
\bibitem{barz} H.~W.~Barz, B.~L.~Friman, J.~Knoll and H.~Schulz,
        Nucl. Phys. A 519 (1990) 831.
\bibitem{Udo} J.~Bolz, U.~Ornik and R.~M.~Weiner, Phys. Rev. C 46 (1992) 2047.
\bibitem{shurehand} E.~V.~Shuryak, Nucl. Phys. A 566 (1994) 559c.
\bibitem{loopies} P.~Rehberg and S.~P.~Klevansky, University of
        Heidelberg Report No.~HD--TVP--95--13 (1995),
        {\tt hep-ph/9510221}.
\bibitem{eddie} E.~V.~Shuryak, Phys. Rep. 61 (1980) 71.
\bibitem{pipi} E.~Quack, P.~Zhuang, Y.~Kalinovsky, S.~P.~Klevansky
        and J.~H\"ufner, Phys. Lett. B 348 (1995) 1.
\bibitem{micki} A.~E.~Dorokhov, M.~K.~Volkov, J.~H\"ufner, S.~P.~Klevansky
        and P.~Rehberg, University of Heidelberg Report No.~HD--TVP--95--12
        (1995).
\bibitem{rost} P.~Rehberg, Yu.~L.~Kalinovsky and D.~Blaschke, University
        of Rostock Report No.~MPG--VT--UR~81/96, in preparation.
\bibitem{deGroot} S.~R.~de~Groot, W.~A.~van~Leeuwen, Ch.~G.~van~Weert,
        {\it Relativistic Kinetic Theory} (North Holland, Amsterdam,
        1980).
\bibitem{Reif} F.~Reif, {\it Fundamentals of Statistical and Thermal
        Physics} (McGraw--Hill, New York, 1965)
\bibitem{Thoma} M.~H.~Thoma, Phys. Lett. B 269 (1991) 144.
\bibitem{MueRaf} J.~Rafelski and B.~M\"uller, Phys. Rev. Lett. 48
        (1982) 1066.
\bibitem{MSML} T.~Matsui, B.~Svetitsky and L.~D.~McLerran, Phys. Rev. D
        34 (1986) 783.
\bibitem{na35} T.~Alber et al., Z. Phys. C 64 (1994) 195.
\bibitem{njlthermo} P.~Zhuang, J.~H\"ufner and S.~P.~Klevansky,
        Nucl. Phys. A 576 (1994) 525.
\bibitem{pengfei} P.~Zhuang, Phys. Rev. C 51 (1995) 2256.
\bibitem{quackhen} E.~Quack and P.~A.~Henning, Phys. Rev. Lett. 75
        (1995) 2811.
\bibitem{neise} L.~Neise, V.~Bunakov and J.~H\"ufner, University of
        Heidelberg Report No.~HD--TVP--94--3 (1994), unpublished.
\bibitem{hosata} A.~Hosoya, M.~Sakagami and M.~Takao, Ann. Phys. (NY)
        154 (1984) 229.
\bibitem{dani} P.~Danielewicz, Ann. Phys. (NY) 152 (1984) 239.
\bibitem{henning} P.~A.~Henning, Phys. Rep. 253 (1995) 235.
\bibitem{zhawi} W.~Zhang and L.~Wilets, Phys. Rev. C 45 (1992) 1900.
\bibitem{wojtek} W.~Florkowski, J.~H\"ufner, S.~P.~Klevansky and
        L.~Neise, Ann. Phys. (NY) 245 (1996) 445.
\bibitem{cusie} R.~Cutler and D.~Sievers, Phys. Rev. D 17 (1978) 196.
\end{references}
\end{document}